\documentclass{aastex63}

\usepackage{mathtools}
\usepackage{multirow}
\graphicspath{{Figures/}}

\submitjournal{ApJS}
\accepted{March 14, 2021}

\shorttitle{F10.7 and F30~cm solar radio flux prediction by Kalman filter}
\shortauthors{Petrova et al.}

\begin{document}

\title{Medium-term predictions of F10.7 and F30 cm solar radio flux with the adaptive Kalman filter}

\correspondingauthor{Tatiana Podladchikova}
\email{t.podladchikova@skoltech.ru}

\author{Elena Petrova}
\affiliation{Skolkovo Institute of Science and Technology, Bolshoy Boulevard 30, bld. 1, Moscow 121205, Russia}

\author{Tatiana Podladchikova}
\affiliation{Skolkovo Institute of Science and Technology, Bolshoy Boulevard 30, bld. 1, Moscow 121205, Russia}

\author{Astrid M. Veronig}
\affiliation{Institute of Physics, University of Graz, Universit{\"a}tsplatz 5, A-8010 Graz, Austria}
\affiliation{Kanzelh{\"o}he Observatory for Solar and Environmental Research, University of Graz, Kanzelh{\"o}he 19, 9521, Treffen, Austria}

\author{Stijn Lemmens}
\affiliation{ESA/ESOC, Robert-Bosch-Str. 5, 64293 Darmstadt, Germany }

\author{Benjamin Bastida Virgili}
\affiliation{IMS GmbH@ESA/ESOC, Darmstadt, Germany}

\author{Tim Flohrer}
\affiliation{ESA/ESOC, Robert-Bosch-Str. 5, 64293 Darmstadt, Germany }

\begin{abstract}
The solar radio flux at F10.7~cm and F30~cm is required by most models characterizing the state of the Earth's upper atmosphere, such as the thermosphere and ionosphere to specify satellite orbits, re-entry services, collision avoidance maneuvers and modeling of space debris evolution. We develop a method called RESONANCE (“Radio Emissions from the Sun: ONline ANalytical Computer-aided Estimator”) for the prediction of the 13-month smoothed monthly mean F10.7 and F30 indices 1-24 months ahead. The prediction algorithm includes three steps. First, we apply a 13-month optimized running mean technique to effectively reduce the noise in the radio flux data. Second, we provide initial predictions of the F10.7 and F30 indices using the McNish-Lincoln method. Finally, we improve these initial predictions by developing an adaptive Kalman filter with the error statistics identification. The root-mean-square-error of predictions with lead times from 1 to 24 months is 5-27~sfu for the F10.7 and 3-16~sfu for F30 index, which statistically outperforms current algorithms in use. The proposed approach based on Kalman filter is universal and can be applied to improve the initial predictions of a process under study provided by any other forecasting method. Furthermore, we present a systematic evaluation of re-entry forecast as an application to test the performance of F10.7 predictions on past ESA re-entry campaigns for payloads, rocket bodies, and space debris that re-entered from June 2006 to June 2019. The test results demonstrate that the predictions obtained by RESONANCE in general also lead to improvements in the forecasts of re-entry epochs.
\end{abstract}


\section{Introduction} \label{sec:intro}
Forecasting of solar activity with different lead times from hours to decades is highly important for many space weather applications \citep{Pesnell2012,Pesnell2016}. In particular, predictions of the solar radio flux at wavelengths of F10.7~cm and F30~cm are required for planning and performing space operations, re-entry predictions \citep{Virgili2017}, collision avoidance \citep{Merz2017}, and orbital lifetime computation. The evolution of the orbital space debris environment is  dependent on the decay rates of the objects over time \citep{Virgili2014}. 

F10.7~cm is the flux density of solar radio emissions at a wavelength 10.7~cm averaged over an hour \citep{Tapping2013}, also called the F10.7 solar radio flux or F10.7 index. It comprises a time-varying mix of different emission mechanisms over the solar disc. Together with the F30 index, it serves as a solar proxy of the ultraviolet (UV) solar emission, which heats the Earth's upper atmosphere \citep{Dudok2014,Vourlidas2018}. In comparison to the true solar UV irradiance measurements, the F10.7 index has long records. The related changes in the thermosphere density cause variations of the atmospheric drag acting on orbiting objects and need to be accurately modeled via atmosphere models. All of them, including the ISO recommended models, DTM\footnote{DTM, the Drag Temperature Model, is a semi-empirical model describing the temperature, density, and composition of the Earth’s thermosphere.} \citep{Bruinsma2015}, NRLMSISE\footnote{NRLMSISE is an empirical, global model of the Earth's atmosphere from ground to space. NRL stands for the US Naval Research Laboratory. MSIS stands for mass spectrometer and incoherent scatter radar, the two primary data sources for development of earlier versions of the model.} \citep{Picone2002}, and GOST\footnote{The GOST-2004 is a Russian Governmental Standard, with code number R 25645.166 – 2004, dedicated to Earth upper atmosphere density model for ballistic support of flights of artificial Earth satellites.} \citep{Gost2004}, considered and assessed by \cite{Virgili2017ISO}, require F10.7 or F30 as a prime for the solar input. As was shown in the study by \cite{Dudok2014}, in application to DTM2012\footnote{An upgrade of the DTM model.}, F30 showed superior performance for thermosphere modeling due to a strong presence of thermal bremsstrahlung in the F30~cm flux, which has a strong counterpart in the UV bands.  

Forecasting of solar activity is relevant on different time scales, which are dominated by different processes. 	\textit{Short-term} forecast is most often used to refer to forecast lead times of several days (up to 30 days; i.e. up to one solar rotation). \textit{Medium-term} forecast is a forecast for a period when a trend may emerge in the dynamics of the process (like due to the 27-day recurrence of sunspots and related activity due to solar rotation), which can be detected and predicted. Medium-term solar activity forecast usually means the predictions over one to several months in advance. \textit{Long-term} forecast is made over a decade or more, i.e. refers to solar cycle time scales. Such forecasts typically deal with the prediction of the very general characteristics of a solar cycle such as its amplitude, peak time and period. Medium and long-term predictions of continuous monthly solar activity indices, which can be quite variable, is quite a delicate task and the predictions can only refer to some trends in the data. Thus, it is also very important first to smooth the data to better detect the trend, which can be predicted.
 
Various reviews of solar activity prediction methods have been published over the last decades \cite[e.g.][]{Hathaway1999, Hathaway2009, Hathaway2010,Pesnell2012,Pesnell2016}, and the most recent and extensive one by \cite{Petrovay2020}. 
Most of these reviews concentrate on sunspot number prediction techniques. The current approaches can be divided, in general, into three groups. The first group comprises so-called precursor methods based on the identification of some properties of the current cycle, which have predictive power for the next cycle. Long-term predictions are generally done using precursors such as geomagnetic activity \citep{OhlOhl1979, Feynman1982, GonzalezSchatten1988, Thompson1993, WilsonHathawayReichmann1998}, polar magnetic fields \citep{SchattenSofia1987, SchattenDynamo1978, Schatten1996, Schatten2005, Svalgaard2005, WangSheeley2009, MunozJaramillo2012}, and sunspot number records or other solar activity indices \citep{Ramaswamy1977, LantosSkewness2006, CameronSchlussler2008, Kane2008, PodladchikovaLefebvreLinden2008, Podladchikova2011, Podladchikova2017}. The second group combines physics-based predictions using surface flux transport models and dynamo theory \citep{Nandy2002, DikpatiGilman2006, CameronSchlussler2007, Choudhuri2007, Henney2012, CameronSchlussler2015}. Finally, the third group represents a variety of statistical methods, which do not require any knowledge of the physics involved \citep{Macpherson1995, Fessant1996, Zhang1996, Conway1998, Sello2001, Aguirre2008, Braja2009, Liu2010, Kakad2020} and can be applied for medium and long-term forecasting. Daily short-term prediction methods of F10.7 and F30 indices use either modifications of (auto-) regression models \citep[e.g.][]{Dmitriev1999,Liu2010,Henney2012} or neural network techniques \citep[e.g.][]{Huang2009,Yaya2017}. At the same time, the manual forecast method that is currently used at SIDC (Solar Influcences Data Center, Belgium) for short-term F10.7 prediction may be more accurate \citep{Devos2014}.
 
Despite the large variety of prediction methods of solar activity discussed above, the operational medium-term prediction of solar radio fluxes (i.e. predictions several months in advance) is available only with SOLMAG, the solar and geomagnetic activity prediction model employed by ESA's Space Debris Office \citep{Mugellesi1993,Virgili2014}. It is based on the well-known regression technique of the medium-term solar cycle prediction, which was introduced by \cite{Mcnish1949}. It uses the averaging over past cycles to form a mean cycle as a first approximation, and the difference between the mean cycle and smoothed sunspot numbers to predict future solar activity. \cite{Stewart1970} introduced a modification of the McNish-Lincoln method to predict the monthly mean values of sunspot numbers. Application of the McNish-Lincoln method to the F10.7 index and geomagnetic indices was presented in \cite{Holland1984, Niehuss1996}. The authors offered to construct the mean cycle by using a Lagrangian linear regression statistical technique. This modification is also used by SOLMAG.
	
\cite{Podladchikova2012} presented one of the latest modifications of the McNish-Lincoln method with the development of an adaptive Kalman filter applied to the prediction of sunspot numbers 1-12 months ahead. The developed algorithm was also implemented to improve the medium-term predictions provided by the Standard and Combined methods.
The Standard Method is based on an interpolation of Waldmeier's standard curves and provides the predictions of the cycle evolution from the average shape of past solar cycles with a specific cycle maximum \citep{Waldmeier1968}. The Combined Method is based on combination of Waldmeier's idea of standard curves and a prediction of the next cycle peak using the aa geomagnetic index  \citep{Denkmayr1997, Hanslmeier1999}. The improved Kalman filter predictions for all three methods are monthly updated at the
SILSO (Sunspot Index and Long-term Solar Observations; http://sidc.be/silso), which is the World Data Center for the production, preservation and dissemination of the international sunspot number.

In this study, we are focusing on a medium-term prediction of F10.7 and F30 indices by developing a new method called RESONANCE (“Radio Emissions from the Sun: ONline ANalytical Computer-aided Estimator”). First we develop a prediction method of smoothed F10.7 and F30 indices from 1 to 24 months ahead. The method is based on the further development of the adaptive Kalman filter, which is applied to improve medium-term predictions originally produced by the McNish-Lincoln method. Second, we evaluate the performance of the newly developed prediction technique and compare its results with the current state-of-the-art SOLMAG model for medium-term F10.7 predictions. Finally, to demonstrate one of the main applications of the solar radio index predictions, we re-evaluate past ESA's re-entry campaigns for 602 payloads and rocket bodies, and 2344 space debris that re-entered from June 2006 to June 2019, using the F10.7 index forecasts from the newly developed method and comparing them with those from the SOLMAG model used by ESA.

\section{Prediction algorithm of the solar radio flux}
We develop a method to predict the F10.7 and F30 indices with lead times of 1 to 24 months. The prediction algorithm includes three main steps. First, to process the short-term fluctuations in the measured radio flux we use a 13-month optimized running mean (described in Section~\ref{subsection_data}). Second, we use the McNish-Lincoln method to provide initial predictions of the F10.7 and F30 indices (described in Section~\ref{subsection_ML}). However, the 13-month running mean of the radio flux, which is used as the input to the McNish-Lincoln method, is available at the price of a 6-month delay with respect to the current time. The radio flux activity over the last 6-month period until the current time is considered to be unknown because of the noise component that is intrinsic to the available monthly mean data. To remove this drawback, in a third step, we develop an adaptive Kalman filter with noise statistics identification, which assimilates the monthly mean radio flux data over the last 6-month period and provides an estimation of the smoothed radio flux at the current time (Section~\ref{subsection_KF}). The reconstructed radio flux at the current time becomes a new starting point for the predictions, which allows the McNish-Lincoln method to perform without a 6-month delay and thus produce an improved forecast. Throughout the text, when we use the phrase ``initial McNish-Lincoln predictions'', we refer to the values directly produced by the McNish-Lincoln method; with ``McNish-Lincoln+Kalman filter predictions (M\&L+KF)'' we mean the improved predicted values produced by applying the combination of the McNish-Lincoln method and the Kalman filter technique (RESONANCE). We also derive analytical expressions for the errors of both the initial McNish-Lincoln predictions and the McNish-Lincoln+Kalman filter predictions. Since those are not yet available in the literature, we include the error derivation in Appendix~\ref{ML_error}.

\subsection{McNish-Lincoln method} \label{subsection_ML}
The McNish-Lincoln method is based on the construction of a mean cycle by employing all past solar cycles, here as characterized by the F10.7 and F30 indices. Cycles are aligned on the month of the activity minimum, and for each time step, the average value is calculated. The predicted solar radio flux $F_{m}$ for the future month $m$ is calculated according to the following equation:
\begin{equation} \label{eq_ML}
F_{m}  =  \bar{F}_{m}+k_{mi}(F_{i} - \bar{F}_{i})  
\end{equation}
Here, the mean value $\bar{F}_m$ serves as a first estimation of the solar radio flux at any future month $m$. This estimation is refined by adding a correction term to it, namely the difference between the last known smoothed value ${F}_i$ at month $i$ and the value of mean cycle $\bar{F}_i$ at month $i$. $k_{mi}$ is a correction coefficient identified from the least squares method, which is represented by Equation~(\ref{eq_A22}) in Appendix~\ref{ML_error}. The prediction error of the McNish-Lincoln method is given by Equation~(\ref{eq_A47}), which we derive in analytical form in Appendix~\ref{ML_error}. 

In the next subsection we describe the procedure of data preparation and construction of a mean cycle required by the McNish-Lincoln method.

\subsection{Data preparation} \label{subsection_data}
For our analysis, we use monthly mean data of the F10.7 index from the Ottawa and Penticton Radio Observatories, and F30 index from the Toyokawa and Nobeyama facilities, which provide the longest and continuous solar radio flux records. The F10.7 and F30 indices are given in solar flux units (sfu) with one 1 sfu=$10^{-22}\ \mathrm{W}\ \mathrm{m}^{-2}\ \mathrm{Hz}^{-1}$. To filter out short-term fluctuations in the data and to isolate the component associated with the long-term behavior of the solar cycle, a 13-month running mean is widely used \citep{Hathaway1999}. This procedure represents a boxcar averaging of the monthly mean data centered on a given month with equal weights for months $-5$ to $+5$ and a half weight for the months at the start $-6$ and at the end $+6$. However, as pointed out by \cite{Hurst1970, Hathaway2010, MunozJaramillo2012, Podladchikova2017} the traditional 13-month running mean does a poor job of filtering out high-frequency variations. Thus, we process the noisy record in radio flux data using a 13-month optimized running mean. The advantages of this method over the traditional 13-month running mean are demonstrated and discussed in \citet{Podladchikova2017}. The technique is based in finding a balance between the fidelity to the data and the smoothness of an approximating curve. The smoothed values are derived from the minimization of the functional
\begin{equation}\label{eq_functional_J}
J = \beta \sum_{i=1}^{n}(\hat{F}_{i}-{F}_{i})^2+\sum_{i=1}^{n-2}(F_{i+2}-2F_{i+1}+F_{i})^2
\end{equation}
Here, $\hat{F}_{i}$ is the monthly mean radio flux, $F_{i}$ is the smoothed flux at month $i$, and $\beta$ is smoothing constant, which determines how closely the approximating curve fits the data. For our analysis we use a smoothing constant of $\beta = 0.01$, 
that, as was shown in \cite{Podladchikova2017}, provides an effective filtration of noise. The rationale behind this approach is to satisfy two intrinsically conflicting criteria that are used to examine the quality of the approximation of measurement data by a smoothed curve. The data fidelity is evaluated by minimizing the sum of the squared deviations between the fit and the data (first term in Equation~(\ref{eq_functional_J})), and the smoothness is evaluated by minimizing the sum of squared second derivatives of the fit curve (second term in Equation~(\ref{eq_functional_J})). The resulting smoothed values minimizing the functional $J$ can be obtained from the solution of a system of $n$ normal equations with $n$ unknowns, where $n$ is the number of available monthly mean data. The technique has been further developed to be applied as the 13-month optimized running mean (see details in \citet{Podladchikova2017}).

\begin{figure}  
	\epsscale{0.8}
	\plotone{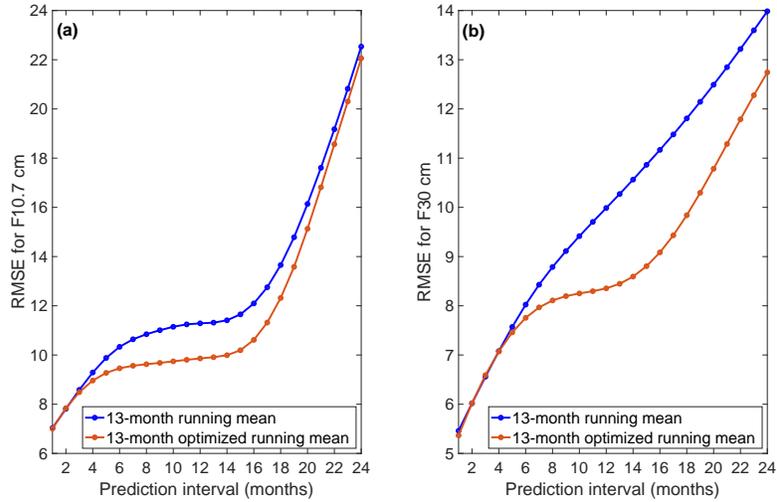}
	\caption{Root-mean-square error (RMSE; in sfu) of the McNish–Lincoln predictions 1-24 months ahead for cycle 24 using the optimized smoothing technique (red line) and the 13-month running mean (blue line). (a) F10.7~cm (b) F30~cm.} \label{fig1}
\end{figure}

Figure~\ref{fig1} shows the root-mean-square-error (RMSE) of the radio flux predictions 1-24 months ahead obtained by the McNish–Lincoln method (with the mean cycle employing cycles 8-23, see below on data preparation for the mean cycle construction).
Panel~(a) shows the RMSE for the F10.7 predictions and panel~(b) for F30. The red line represents the errors for the 13-month optimized running mean and blue line for the traditional 13-month running mean. As can be seen from Figure~\ref{fig1}, the application of the 13-month optimized running mean reduces the errors of radio flux predictions compared to the traditional 13-month running mean, especially on longer prediction intervals. The maximal improvements of prediction accuracy with the 13-month optimized running mean reach 13\% for 10-month forecast of F10.7 and 19\% for 14-month forecast of F30 in cycle 24.

Data for both F10.7 and F30 indices are available since November 1951, which constitutes only six 11-year solar cycles 19-24 and does thus not allow constructing the mean cycle employing all past cycles. However, this dataset can be augmented by reconstructing the radio flux back to the first cycle using the smoothed sunspot numbers since the correlation between these times series is very high. To reconstruct the solar radio flux data, \cite{Virgili2014} used a quadratic fit for SOLMAG -- solar and geomagnetic activity prediction model developed by ESA. We use a third-order linear regression and obtain the following relationship between the smoothed radio flux data $F_{i}$ and the smoothed sunspot numbers $R_{i}$ for each month $i$: 
\begin{equation}\label{eq_regression_flux_sunspot}
F_{i} = \beta_{0}+\beta_{1}R_{i}+\beta_{2}R_{i}^2+\beta_{3}R_{i}^3	
\end{equation}
Here, the vector of regression coefficients $\beta^{F10.7}=\left|66.1404,\ 0.4572,\ 0.0018, \-4.4602\cdot10^{-6} \right|^{T}$ for F10.7 and $\beta^{F30}=\left|41.3547,\ 0.3669,\ 7.6089\cdot10^{-4}, \ -2.5785\cdot10^{-6} \right|^{T}$ for F30 are determined from the least squares method. 

\begin{figure}  
	\plotone{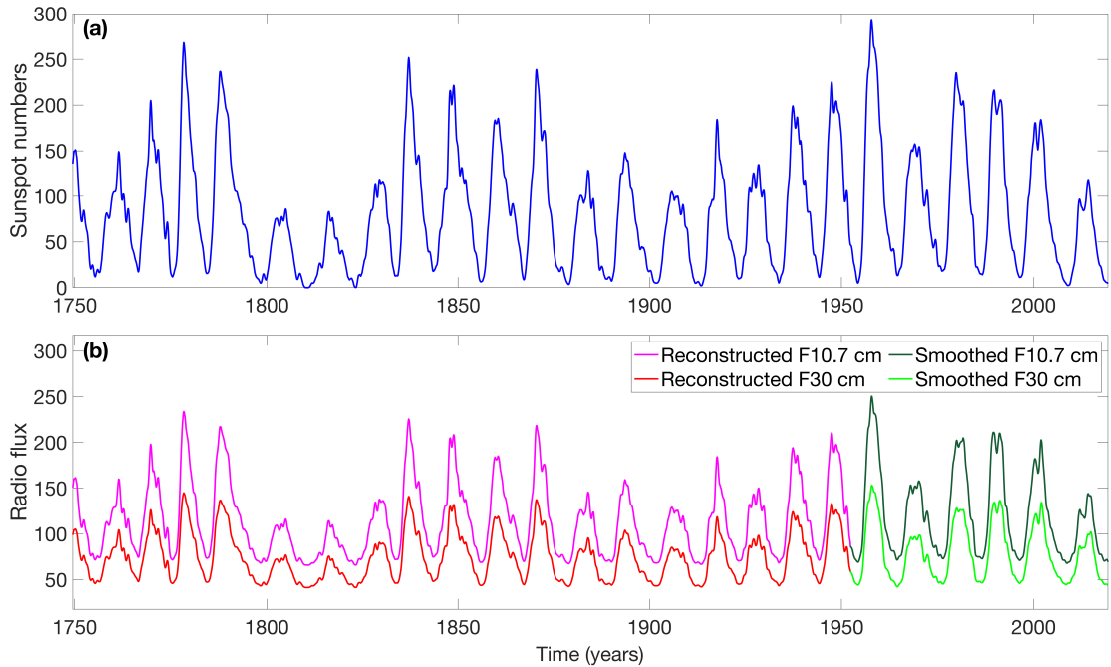}
	\caption{(a) Smoothed monthly sunspot numbers used for the radio flux reconstruction. (b) Smoothed F10.7 (dark green line) and F30 (light green line) based on the measured data, and the reconstructed F10.7 (magenta line) and F30 (red line) back to cycle 1.}	\label{fig2}
\end{figure}

The F10.7 and F30 indices reconstructed on the basis of Equation~(\ref{eq_regression_flux_sunspot}) together with the smoothed sunspot numbers are shown in Figure~\ref{fig2}. The monthly sunspot numbers are from SILSO, including the corrections applied in \citet{Clette2016}. The dark and light green lines in panel~(b) show the smoothed F10.7 and F30 indices based on the measured data, while magenta and red lines represent the reconstructed data using the smoothed sunspot numbers given in panel~(a). The correlation coefficient between the measured and reconstructed F10.7~cm flux is 0.99 and the standard deviation is 5.43 sfu for cycles 19-24. For the F30~cm flux the correlation coefficient is 0.98 and the standard deviation is 5.71 sfu.
As can be seen from Figure~\ref{fig2}, the dynamics of the solar radio flux reveals a wide variety for the different solar cycles in terms of amplitude, duration, shape, steepness of the rise, etc. This variability is taken into account by the construction of the mean cycle. However, an important question to answer is whether to use all the reconstructed cycles till the first one or to reject early measurements since, according to \cite{Mcnish1949}, cycles prior to no.8 are subject to a high level of uncertainty due to non-consistent observations. Thus, the inclusion of a different number of cycles influences the shape of the mean cycle used in the McNish-Lincoln method. As a consequence, rejection of a certain number of cycles may improve the accuracy of the prediction for particular cycles, while worsening for others. Additional uncertainty is related to the errors in the reconstruction of the solar radio flux from the sunspot numbers using Equation~(\ref{eq_regression_flux_sunspot}). As shown in Figure~\ref{fig2}, the differences in the shape of the measured solar radio flux (dark and light green lines) and sunspot numbers (blue line) are observed in particular during the periods of solar maxima for cycles 21-23. For instance, while the second peak of sunspot cycle 22 is smaller than the first one, the amplitudes of both peaks are comparable for the radio flux data. For cycle 23, we observe almost similar peaks in the sunspot data, whereas in the radio flux data the first peak is significantly smaller than the second one. In light of this, in Section~\ref{Results_ML_KF}, we discuss the performance of the proposed algorithm for different mean cycles employing cycles 8-24 and only the ones where radio measurements are available (cycles 19-24).
 
\subsection{Improvement of McNish-Lincoln predictions with an adaptive Kalman filter} \label{subsection_KF}
\begin{figure}  
    \epsscale{0.8}
	\plotone{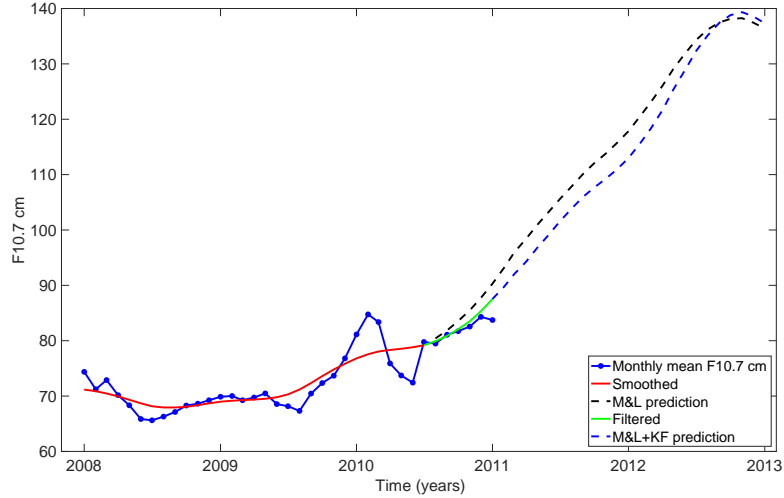}
	\caption{Illustration of the concept to use an adaptive Kalman filter for the improvement of the initial McNish-Lincoln predictions. The blue dotted line gives the monthly mean F10.7. The solid red line shows the smoothed monthly F10.7. The dashed black line shows the initial McNish-Lincoln (M\&L) predictions 30 months ahead, where the first 6 months represent the estimated radio flux dynamics till the current time, and the following 24 months give the predictions into the future. The solid green line shows the reconstructed smoothed F10.7 till the current time using the Kalman filter (KF). The estimated smoothed radio flux at the current time is used as a new starting point for the improved McNish-Lincoln prediction (dashed blue line, M\&L+KF).}
	\label{fig3}
\end{figure}

The McNish-Lincoln prediction uses the latest available smoothed radio flux at the price of a 6-month delay with respect to the current time. Figure~\ref{fig3} illustrates an example, when the last 13-month running mean is available for July 2010 (solid red line), and this value with the 6-month delay compared to the current time (January 2011) becomes a starting point for the McNish-Lincoln predictions (dashed black line). The method does not directly use the last available six monthly mean flux data (solid blue line) because of the stochastic component intrinsic to them, though they give significant information about the radio flux evolution in the future. We propose to improve the initial McNish-Lincoln predictions by taking into account the dynamics of the monthly mean radio flux over the last six months. To achieve this goal, we develop an adaptive Kalman filter with noise statistics identification to reconstruct the smoothed radio flux curve up to the current time (solid green line). 
The estimated radio flux at the current time becomes a new starting point for the improved McNish-Lincoln prediction (dashed blue line), which now performs without a 6-month delay. 

\subsubsection{Model description}\label{subsubsection_State_Space}
The Kalman filter requires that an investigated dynamical system can be described by a mathematical model in state-space representation, which reflects the changes in the system states. We present the state-space model of the solar radio flux activity in the following way
\begin{equation}\label{eq_state_equation}
F_{i} =\Phi_{i,i-1}F_{i-1}+w_{i}, \qquad (i=1,2,\ldots,6)
\end{equation}
Here, $F_{i}$ is a state vector of the smoothed solar radio flux to be estimated over the last 6-month period until the current time. The transition state matrix $\Phi_{i,i-1}$ relates a current state vector $F_{i-1}$ to a one-step-ahead predicted state vector $F_{i}$. The determination of the matrix $\Phi_{i,i-1}$ usually comes from a physical model of the process under study, which in our case is not specified. However, to provide the best estimation of the unknown solar radio flux $F_{i}$ the identification of matrix $\Phi_{i,i-1}$ is crucially important. To get the estimates of the matrix $\Phi_{i,i-1}$, we suggest using the initial McNish-Lincoln predictions $\hat{F}_{i}^{init}$, that reflect the evolution of the solar radio flux in the future:
\begin{equation}\label{transition_matrix_phi_all}
\Phi_{i,i-1} =\frac{\hat{F}_{i}^{init}}{\hat{F}_{i-1}^{init}}, \qquad (i=2,3,\ldots,6)
\end{equation}
The transition state matrix that relates the last available 13-month running mean $F_{0}$ to
the one-month-ahead prediction $\hat{F}_{1}^{init}$ is estimated as 
\begin{equation}\label{transition_matrix_phi_init}
\Phi_{1,0} =\frac{\hat{F}_{1}^{init}}{F_{0}}
\end{equation}
However, the transition state matrix $\Phi_{i,i-1}$ constructed in this way may have unknown errors related to the  
deviations of the initial McNish-Lincoln predictions $\hat{F}_{i}^{init}$ from the 
smoothed radio flux activity $F_{i}$. The presence of model errors arising from the imperfect description of the dynamics of the process can cause essential estimation errors. Therefore, we add to the state Equation~(\ref{eq_state_equation})
the uncorrelated and unbiased model noise $w_{i}$ with unknown variance $\sigma_{w_{i}}^2$, that describes the random errors of the initial McNish-Lincoln predictions.

Let $F_{i}^{m},\ (i=1,2,\ldots,6)$ denote the monthly mean radio flux data available over the last last 6-month period up to the current time. To relate the available measurements $F_{i}^{m}$ to the state vector $F_{i}$ we need to introduce the following measurement equation
\begin{equation}\label{eq_measurement_equation}
F_{i}^{m} =F_{i}+\eta_{i}, \qquad (i=1,2,\ldots,6)
\end{equation}
Here, $\eta_{i}$ is an uncorrelated and unbiased measurement noise with unknown variance $\sigma_{\eta_{i}}^2$, that describes the random deviations of monthly mean measurements $F_{i}^{m}$ from the smoothed flux $F_{i}$. 

Equations~(\ref{eq_state_equation})~and~(\ref{eq_measurement_equation}) represent a stochastic state-space model of the solar radio flux activity based on the initial McNish-Lincoln predictions. However, the statistical characteristics 
of model and measurement noise $w_{i}$ and $\eta_{i}$ introduced into the model are unknown.
The application of a Kalman filter provides an optimal estimation with respect to the minimization of the covariance matrix of the estimation error independent of the distribution of noise \citep{Kalman1960,Sage1971}. However, the correct specification of model error statistics is essentially important since uncertainties in the model and measurements may cause large errors in the prediction and lead to failure of a Kalman filter algorithm. Therefore, to provide the best estimation of the smoothed radio flux $F_{i}$, we identify the unknown variances $\sigma_{w_{i}}^2$ and $\sigma_{\eta_{i}}^2$ of the model and measurement noise $w_{i}$ and $\eta_{i}$ using the consistent noise statistics identification method presented in \cite{Podladchikova2012}. According to \cite{Hathaway1994}, the variance of statistical errors in monthly mean sunspot numbers is proportional to their values. We use the same assumption about the radio flux dynamics and consider the variances $\sigma_{w_{i}}^2$ and $\sigma_{\eta_{i}}^2$ to be proportional to the smoothed radio flux $F_{i}$
\begin{equation}\label{eq_coef_noise}
\begin{array}{c} 
\sigma_{w_{i}}^{2} =\alpha_{w}F_{i}\\
\sigma_{\eta_{i}}^{2} =\alpha_{\eta}F_{i}
\end{array}
\end{equation}
Here, $\alpha_{w}=0.2$ and $\alpha_{\eta}=2.6$ are the identified coefficients of proportionality.
Note, that the application of a Kalman filter provides 
an optimal estimation with respect to the minimization of the 
covariance matrix of the estimation error independent 
of the distribution of noise (Kalman 1960; Sage 1971).
However, we need to know the variances of the model and measurement noise,
which we identified taking into account their increase with the increase of the radio flux activity. 
The obtained estimates of the \textit{variable} variances $\sigma_{w_{i}}^{2}$ and $\sigma_{\eta_{i}}^{2}$, which take into account the heteroscedasticity of the radio flux data, are further incorporated into the Kalman filter algorithm.

\subsubsection{Kalman filter algorithm} \label{subsubsection_reccurent_KF}
We estimate the smoothed radio flux described by the state-space model 
(Equations~(\ref{eq_state_equation})~and~(\ref{eq_measurement_equation})) 
using a Kalman filter, that represents a very powerful and efficient data assimilation technique 
to estimate the state of a process \cite{Kalman1960}. The Kalman filter is a recurrent algorithm which performs in a two-step process: 

\textit{Prediction} is performed to estimate the future state vector of smoothed radio flux $F_{i}$ one step ahead 
\begin{equation}\label{KF_extrapolation}
\hat{F}_{i,i-1} = \Phi_{i,i-1}\hat{F}_{i-1,i-1}, \qquad (i=1,2,\ldots,6)
\end{equation}
Here, $\hat{F}_{i,i-1}$ represents a predicted estimate of state ${F}_{i}$ at step $i$. We use the first subscript $i$ to indicate the time on which the estimation of the smoothed radio flux is made and the second subscript $i-1$ to denote the number of monthly mean radio flux observations $F_{1}^{m},\ F_{2}^{m}, \ldots,F_{i-1}^{m}$ assimilated by the algorithm to make the prediction. $\hat{F}_{i-1,i-1}$ represents a filtered estimate of state ${F}_{i-1}$ at the previous time-step $i-1$.  Here, the second subscript $i-1$ denotes that the filtered estimate $\hat{F}_{i-1,i-1}$ is obtained on the basis of monthly mean radio flux observations $F_{1}^{m},\ F_{2}^{m}, \ldots,F_{i}^{m}$. 

The prediction accuracy of $\hat{F}_{i,i-1}$ in the algorithm is characterized by the variance of prediction error $\sigma_{i,i-1}^2$ in the following way
\begin{equation}\label{KF_extrapolation_error}
\sigma_{i,i-1}^2 = \Phi_{i,i-1}^{2}\sigma_{i-1,i-1}^2+\sigma_{w_{i}}^{2}, \qquad (i=1,2,\ldots,6)
\end{equation}
Here, $\sigma_{i-1,i-1}^2$ is the variance of filtration error at step $i-1$. 
To initiate the Kalman filter procedure, we use the last available value of monthly mean radio flux at the current time as an estimate of the initial state $\hat{F}_{0,0}$ with a variance of the prediction error $\sigma_{0,0}^2=0$.

2. \textit{Filtration} is made to incorporate a new measurement $F_{i}^{m}$ for obtaining an improved estimate of state ${F}_{i}$ at step $i$:
\begin{equation}\label{KF_filtration}
\hat{F}_{i,i} = \hat{F}_{i,i-1} + K_{i}(F_{i}^{m}-\hat{F}_{i,i-1}), \qquad (i=1,2,\ldots,6)
\end{equation}
The estimation accuracy of the filtered value $\hat{F}_{i,i}$ is characterized by the variance of the filtration error $\sigma_{i-1,i-1}^2$
\begin{equation}\label{KF_filtration_error}
\sigma_{i,i}^2 = (1-K_{i})\sigma_{i,i-1}^2 
\end{equation}
Here, $K_{i}$ is the filter gain responsible for the relative weight given to the current measurement $F_{i}^{m}$ 
and the predicted state $\hat{F}_{i,i-1}$ defined as
\begin{equation}\label{KF_filter_gain}
	K_{i} = \frac{\sigma_{i,i-1}^{2}}{\sigma_{i,i-1}^{2}+\sigma_{\eta_{i}}^{2}}
\end{equation}
The unknown variances of model and measurement noise $\sigma_{w_{i}}^{2}$ and $\sigma_{\eta_{i}}^{2}$ in Equations~(\ref{KF_extrapolation_error})~and~(\ref{KF_filter_gain}) are identified from Equation~(\ref{eq_coef_noise}), where a filtered estimate $\hat{F}_{i-1,i-1}$ at step $i-1$ is used instead of the unknown radio flux $F_{i}$. 

The developed adaptive Kalman filter algorithm assimilates the monthly mean radio flux data over the last 6-month period and, as a final output, produces the filtered estimate $\hat{F}_{6,6}$, which represents an estimate of the unknown smoothed radio flux $F_{i}$ at the current time. The root-mean-square errors (``RMSE'') of this estimate for cycles 19-24 for both F10.7 and F30 indices are shown in Table~\ref{table1} (the row ``M\&L+KF''). To get the estimate of the smoothed radio flux $F_{i}$ at the current time with the McNish-Lincoln method, we need to produce a 6-month lead forecast, as the last 13-month running mean, which is used as the input to the McNish-Lincoln method, is available only with a 6-month delay with respect to the current time. The corresponding RMSE of the McNish-Lincoln predictions are shown in the row ``M\&L''. The column ``$\Delta$RMSE'' shows the advantages (in percent) of estimating the smoothed solar radio flux using the proposed Kalman filter compared to the initial 6-month forecast with the McNish-Lincoln method. As shown in Table~\ref{table1}, the developed adaptive Kalman filter reduces the estimation errors of the smoothed solar radio flux at the current time by 23-46\% for F10.7~cm and by 29-49\% for F30~cm. onthly mean data. 
\begin{table}
	\centering
	\caption{The root-mean-square errors (``RMSE''; in units of sfu) of the smoothed solar radio flux estimation at the current time, which is considered to be unknown, as the last 13-month running mean is available only with a 6-month delay with respect to the current month. The row ``M\&L'' shows the RMSE of McNish-Lincoln method, for which the estimation of the smoothed radio flux at the current time is obtained by producing a  6-month lead forecast, as the last 13-month running mean, which is used as the input to the McNish-Lincoln method, is available only with a 6-month delay with respect to the current time. The row `M\&L+KF'' gives the RMSE of the Kalman filter algorithm, which provides the estimation of the smoothed solar radio flux by assimilating the monthly mean radio flux data over the last 6-month period. The column ``$\Delta$RMSE'' shows the improvements (in percent) of
	estimating the smoothed solar radio flux using the Kalman filter compared to the initial 6-month forecast with the McNish-Lincoln method.}
	\label{table1}
	\begin{tabular}{clcccc}
		\hline
		Cycle no. & \multicolumn{1}{c}{Prediction method} & \multicolumn{2}{c}{F10.7~cm} & \multicolumn{2}{c}{F30~cm} \\ 
		& \multicolumn{1}{c}{} & RMSE  & $\Delta$RMSE & RMSE & $\Delta$RMSE \\ \hline
		\multirow{2}{*}{19} & M\&L & 8.11 &  & 5.69 &  \\ 
		& M\&L+KF & 5.14 & 45\%  & 3.14 & 45\% \\ \hline
		\multirow{2}{*}{20} & M\&L & 7.93 &  & 4.95 &  \\ 
		& M\&L+KF & 4.25 & 46\%  & 2.53 & 49\%  \\ \hline
		\multirow{2}{*}{21} & M\&L & 6.99 &  & 3.65 &  \\ 
		& M\&L+KF & 4.86 & 30\%  & 2.6 & 29\%  \\ \hline
		\multirow{2}{*}{22} & M\&L & 13.49 &  & 7.64 &  \\  
		& M\&L+KF & 7.56 & 44\%  & 4.22 & 45\%  \\ \hline
		\multirow{2}{*}{23} & M\&L & 9.18 &  & 6.39 &  \\ 
		& M\&L+KF & 5.03 & 45\%  & 3.32 & 48\%  \\ \hline
		\multirow{2}{*}{24} & M\&L & 6.82 &  & 5.1 &  \\ 
		& M\&L+KF & 5.22 & 23\%  & 3.1 & 39\%  \\ \hline
	\end{tabular}
\end{table}

\section{Results}\label{Results}
\subsection{Forecast performance of M$\&$L+KF compared to the initial McNish-Lincoln predictions} \label{Results_ML_KF}
The estimation of the smoothed radio flux at the current time using the developed adaptive Kalman filter allows us to 
effectively remove an essential drawback of initial McNish-Lincoln predictions, when the last 13-month running mean is available only with a 6-month delay with respect to the current time. To produce an improved radio flux forecast 1-24 months ahead, we again apply the McNish-Lincoln method but use the estimated smoothed radio flux at the current time as the starting point for the predictions. We further refer to it as McNish-Lincoln+Kalman filter predictions. The prediction error of the McNish-Lincoln+Kalman filter is represented by Equation~(\ref{eq_A51}), and its analytical derivation is given in Appendix~\ref{ML_error}. 

Figure~\ref{fig4} shows the root-mean-square error (RMSE) of the radio flux prediction 1-24 months ahead with the McNish-Lincoln+Kalman filter for two different mean cycles used in the McNish-Lincoln method. The red lines give the RMSE for the mean cycle estimated from cycles 8-24, while the blue lines indicate the prediction errors for the mean cycle derived only from the cycles for which radio flux measurements are available (19-24). Note, that each cycle that we predict with the developed algorithm is excluded from the construction of the mean cycle. We show the initial McNish-Lincoln predictions with the dashed lines (M\&L) and the improved ones using the adaptive Kalman filter with the solid lines (M\&L+KF). The two left rows present the RMSE for F10.7, while the two right rows show the prediction errors for F30. 
\begin{figure}  
	\plotone{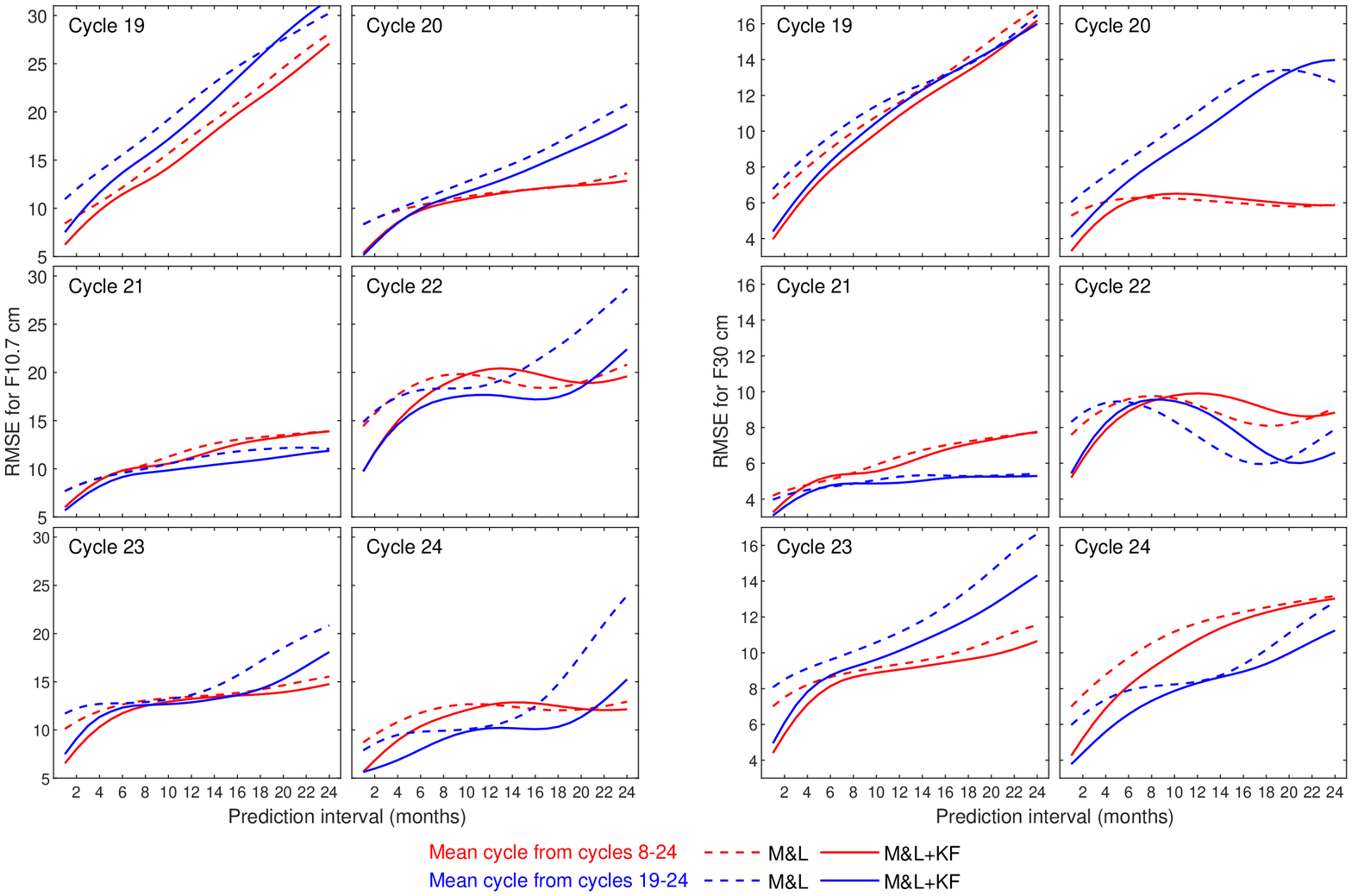}
	\caption{Root-mean-square error (RMSE; in sfu) of the radio flux prediction 1-24 months ahead for two different mean cycles 
		     in the McNish-Lincoln method as function of prediction lead time. The red lines presents the errors for the mean cycle derived from cycles 8-24 and the blue lines shows the RMSE for the mean cycle estimated only from the measured cycles (cycles 19-24). The dashed lines show the initial McNish-Lincoln predictions (M\&L), while the solid lines indicate the improved ones using the adaptive Kalman filter (M\&L+KF). The two left rows present the RMSE for F10.7, while the two right rows present the prediction errors for F30.}
	\label{fig4}
\end{figure}

As can be seen from Figure~\ref{fig4}, in most of the cases, the application of the McNish-Lincoln + Kalman filter (M\&L+KF, solid lines) gives a higher prediction accuracy compared to the initial McNish-Lincoln predictions (M\&L, dashed lines). 
In general, for both methods the prediction error increases with the increase of the prediction lead time. However, while for cycles 22, 23, and 24 (M\&L, dashed blue lines) the errors of the initial McNish-Lincoln predictions of F10.7 cm 12-24 months ahead increase steeply, the application of the McNish-Lincoln + Kalman filter (M\&L+KF, blue solid lines) provides a significantly higher prediction accuracy. The estimation of the radio flux activity over the last 6-month period until the current time with the adaptive Kalman filter allows us to reduce the uncertainties in the dynamics and thus to increase the prediction accuracy. Note, that the different shape of the RMSE as function of lead time in solar cycle 19 (steep, almost linear increase) compared to the other cycles is probably related to the early radio observations that had less stability.

\begin{figure}  
	\plotone{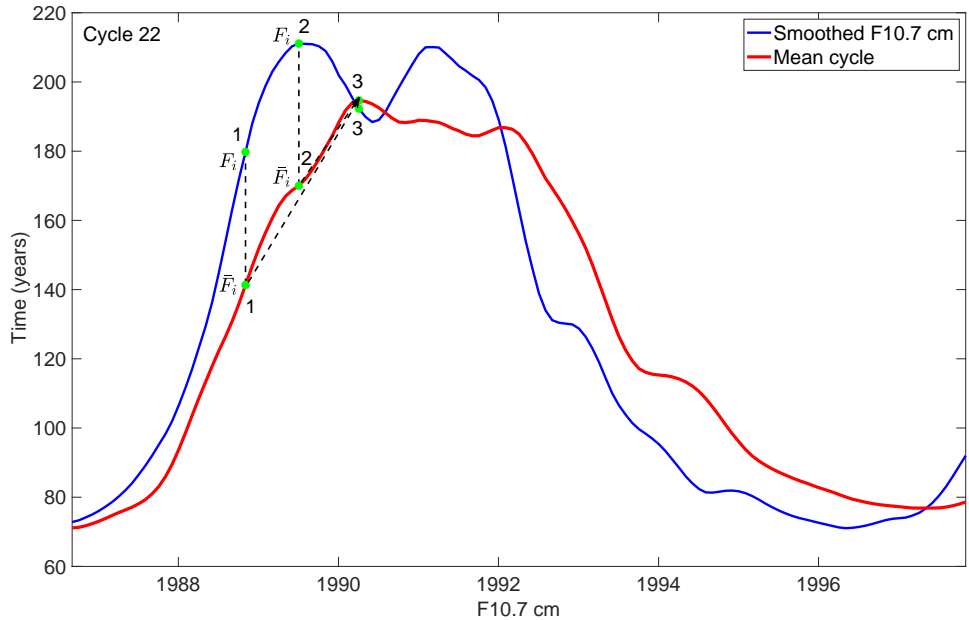}
	\caption{Mean cycle constructed from cycles 8-24 (red line) in comparison with F10.7~cm in cycle 22 (blue line). Green point 3 shows the radio flux to be predicted during the Gnevyshev gap. Green points 1 and 2 give the starting time for prediction 17 and 9 months ahead, respectively.}
	\label{fig5}
\end{figure}

In some cases, for example, in cycle 22, the F10.7 predictions 11-19 months ahead (red lines) and F30 predictions 7-19 months ahead (blue lines) show a better performance with the initial McNish-Lincoln predictions (M\&L, dashed lines). We point out that the indicated prediction interval exhibits a decrease of the prediction errors with increasing lead time, which is untypical for forecasting algorithms. A particularity of cycle 22 is that it has two maxima of almost the same value (Figure~\ref{fig5}, blue line) separated by a significant Gnevyshev gap \citep{Gnevyshev1967, Norton2010}. Also, it has the shortest rise from minimum to maximum of any recorded cycle (just 34 months), and after the first peak, we observe a sharp decrease of the radio flux activity. This specific behavior causes a large mismatch between the dynamics of cycle 22 and the shape of the mean cycle (Figure~\ref{fig5}, red line) that reaches the maximum right at the Gnevyshev gap. When we predict the radio flux near the interval of the Gnevyshev gap (green point 3) starting from the time moment indicated by the green point 2 (9-month ahead prediction), the underestimated value of the mean cycle $\bar{F}_{i}$ leads to an overestimation of the difference $(F_{i} - \bar{F}_{i})$ in Equation~(\ref{eq_ML}), and thus to an overestimation of the prediction. The application of a Kalman filter, which assimilates the monthly mean radio flux values over last 6 months may additionally increase the prediction error. The monthly mean data may indicate a future increase of radio flux activity, while it actually decreases due to a Gnevyshev gap. However, if we start the prediction from the time moment marked by green point 1 (17-month ahead prediction), the difference $(F_{i} - \bar{F}_{i})$ is less distorted due to a smaller mismatch between the shape of the mean cycle and the actual activity, which results in a more accurate prediction for larger prediction intervals. A similar non-typical decrease of prediction errors with increasing prediction time is also observed for the prediction of F10.7 in cycle 24 (Figure~\ref{fig5}, red lines). However, this is not the case for the F30 prediction, as the Gnevyshev gap in cycle 24 for F30 is less prominent compared to F10.7. 

As can be seen from Figure~\ref{fig4}, the accuracy of the predictions depends on both the choice of the mean cycle and the peculiarities of a predicted cycle. The mean cycle estimated from cycles 19-24 has the advantage of choosing available measurements only, but less diversity of the cycles considered (e.g., double-peak solar cycles, long-lasting solar minimum, abrupt rises of solar activity, etc). The mean cycle constructed from cycles 8-24 gives a better representation of the different cycle shapes and possible amplitudes. In general, the mean cycle constructed from cycles 8-24 provides a higher  F10.7 and F30 prediction accuracy for cycles 19, 20, and 23 (red lines), while the mean cycle estimated from the measured cycles (19-24) shows better performance for cycles 21, 22, and 24 (blue lines). On average, the prediction accuracy is 10\% higher for F10.7 and 5\% higher for F30 in case of constructing the mean cycle from cycles 8-24, than only from thee measured ones (19-24). The improvements in prediction accuracy with the application of the McNish-Lincoln + Kalman filter reach 36\% for F10.7 and 39\% for F30. In addition to the mean cycle estimated from cycles 8-24 and 19-24, we also tested the performance of the prediction algorithm for the mean cycle derived from all the available cycles 1-24, cycles 12-24, and cycles 14-24. As the final solution, we chose the mean cycle constructed from cycles 8-24, as, on average, it provides better performance of the McNish-Lincoln predictions compared to other options. 

\begin{figure}
	\plotone{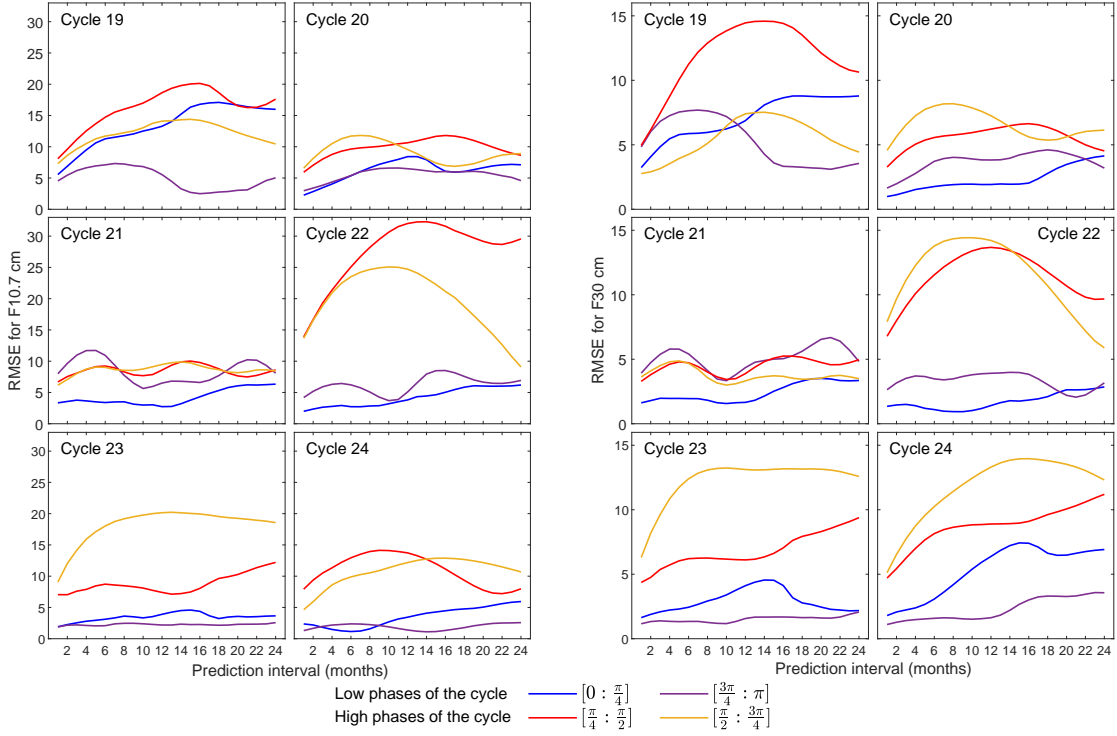}
	\caption{Root-mean-square error (RMSE; in sfu) of the M\&L+KF radio flux prediction 1-24 months ahead for different phases of the cycle. The left panel shows the RMSE for F10.7 and the right panels for F30. The blue and magenta line give the RMSE for low phases of the cycle corresponding to $[0:\frac{\pi}{4}]$ and $[\frac{3\pi}{4}:\pi$] argument of the fitted sine function. The red and yellow line give RMSE for the high phases of the cycle corresponding to $[\frac{\pi}{4}:\frac{\pi}{2}]$ and $[\frac{\pi}{2}:\frac{3\pi}{4}]$ intervals.}
	\label{fig6}
\end{figure}

To investigate the accuracy of predictions for different phases of a cycle instead of dividing the cycle into 2 phases by the point of the cycle maximum, ascending and descending, we followed the procedure: fit each cycle with a sine function with argument $[0:\pi]$ and then divide the cycle into 4 phases for $[0:\frac{\pi}{4}]$, $[\frac{\pi}{4}:\frac{\pi}{2}]$, $[\frac{\pi}{2}:\frac{3\pi}{4}]$,$[\frac{3\pi}{4}:\pi$] intervals of the argument. 
Figure~\ref{fig6} shows the root-mean-square error of predictions provided by the M\&L+KF method with the mean cycle derived from cycles 8-24 for the 4 indicated phases of cycles 21-24. The left panel gives the RMSE for F10.7, while the right panels shows the same for F30. The blue and magenta line give the RMSE for low phases of the cycle corresponding to $[0:\frac{\pi}{4}]$ and $[\frac{3\pi}{4}:\pi$] intervals. The red and yellow lines show the RMSE for the high phases of the cycle, i.e. the $[\frac{\pi}{4}:\frac{\pi}{2}]$ and $[\frac{\pi}{2}:\frac{3\pi}{4}]$ intervals.
As can be seen from Figure~\ref{fig6}, in general the RMSE values are significantly lower in phases of low than high solar activity. Some exceptions exist for cycles 19, 20, and 21, which are related to their shapes and closeness of the actual radio flux activity to the mean cycle.

\subsection{Forecast performance of M$\&$L+KF compared to SOLMAG predictions} \label{Results_ML_SOLMAG}
SOLMAG is a solar and geomagnetic activity prediction model employed by ESA's Space Debris Office. 
It has short, medium, and long-term prediction algorithms, where the last two are also based on the McNish-Lincoln mehod augmented with corrections from \cite{Holland1984}. 
Cycles before 19 are reconstructed based on the sunspot numbers using a quadratic fit \citep{Virgili2014}.  

\begin{figure}
    \epsscale{0.7}
    \plotone{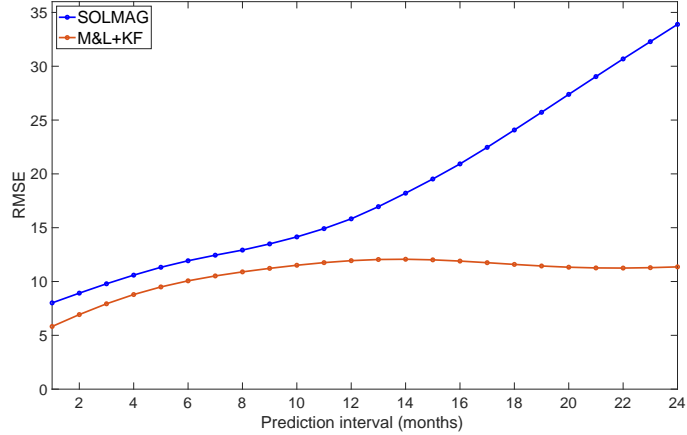}
	\caption{Root-mean-square error (RMSE) of F10.7 prediction 1-24 months ahead for cycle 24. The blue line shows the RMSE for the ESA's SOLMAG method, while the red line for the McNish-Lincoln + Kalman filter (M\&L+KF).}
	\label{fig7}
\end{figure}

Figure~\ref{fig7} shows the performance of the developed M\&L+KF method in comparison with the SOLMAG method for cycle 24. The red line shows the root-mean-square error (RMSE) of F10.7 predictions for M\&L+KF, while the blue line gives the error for SOLMAG. Figure~\ref{fig7} indicates the overall superior performance of  M\&L+KF method for all the prediction intervals. The developed approach statistically outperforms the SOLMAG method by 15.5-66.5\% for cycle 24 with a significant enhancement for longer prediction intervals.
%

\begin{figure} 
    \plotone{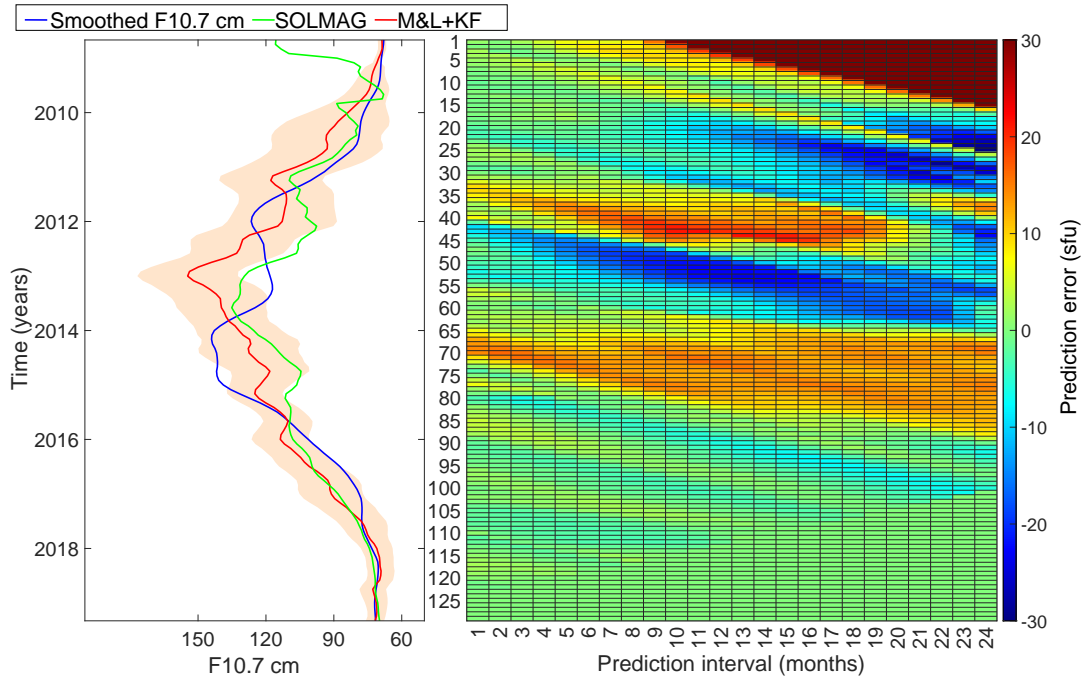}
	\caption{Heat map of the F10.7 predictions by ESA's SOLMAG method and the M\&L+KF method for cycle 24. The left y-axis of the heat map (right panel) shows the month of the predicted radio-flux value starting with September 2008. The x-axis represents the prediction interval (1-24 months ahead). A cell in the heat map gives the absolute F10.7 prediction error for every month of the cycle. Negative prediction errors indicate better performance of the SOLMAG method. Positive prediction errors indicate better performance of the developed M\&L+KF. Left panel shows the evolution of the 10.7 index in cycle 24 (blue line) and an example of the F10.7 predictions 12 months ahead. The green line gives the predictions by SOLMAG method and the red line by the M\&L+KF method. The shaded region represents the $1\sigma$ error (derived with Equation~(\ref{eq_A51}) for the M\&L+KF predictions).}
	\label{fig8}
\end{figure}

To show a more detailed comparison of the performance that takes into account the dependence of the prediction accuracy on the phase of the cycle and the prediction interval, in Figure~\ref{fig8}, we present a so-called heat map (right panel). The left Y-axis of the heat map indicates the month of the predicted F10.7 value, while the X-axis shows how early the forecast is made (1-24 months in advance). A cell in the heat map shows the prediction error of F10.7 for every month of the cycle. In the left panel we plot the time evolution of the 13-month smoothed monthly mean F10.7 index (blue line) together with the 12-month ahead predictions by M\&L+KF (red line) and SOLMAG (green line). The shaded area gives the $1\sigma$ error, which is derived with Equation~(\ref{eq_A51}) for the M\&L+KF predictions, and can be used to provide an uncertainty range for the real-time predictions. In the heat map, we calculate the prediction error as the absolute difference between the predicted and actual radio flux. Colors corresponding to negative values (blue-turquoise) indicate better performance of the SOLMAG method, whereas colors of positive values (red-yellow) indicate better performance of the developed M\&L+KF method. 
As can be seen from the heat map (Figure~\ref{fig8}), the two cycle peaks are in general better predicted with the proposed McNish-Lincoln + Kalman filter. The number of positive cells is around three times larger for the first peak and around five times larger for the second peak than that of the negative ones. However, the prediction accuracy of the Gnevyshev gap interval is mostly better for the SOLMAG method with the number of negative cells six times larger than that of the positive ones. The number of positive and negative cells for the ascending and descending phases is comparable for both prediction methods. However, at the early ascending phase M\&L+KF clearly outperforms SOLMAG predictions, which show difficulties at the very beginning of a solar cycle.

As can be seen from the left panel of Figure 8, M\&L+KF (red line) gives overestimated predictions in the Gnevyshev gap interval due to the reasons discussed in Section~\ref{Results_ML_KF}. At the same time, SOLMAG (green line) gives underestimated predictions of the decreasing radio flux after the first cycle peak. Incorporating different numbers of past cycles into the calculation of the mean cycle changes its shape and affects the accuracy of the predictions. For M\&L+KF we use all the cycles beginning from cycle 8, while the mean cycle in the SOLMAG method employs all the cycles starting from cycle 12.

\section{Tests on past re-entry campaigns}
In this section, we present a systematic evaluation of re-entry forecast and test the performance of the F10.7~cm predictions on past ESA re-entry campaigns for 602 payloads and rocket bodies, and 2344 objects of space debris that re-entered from June 2006 to June 2019 over the full solar cycle. An example of the re-entry prediction campaign for AVUM, the upper stage of a VEGA-01 rocket, coordinated by ESA is presented in \cite{Virgili2017}.

Re-entry is the return of an object from outer space into and through the gases of the atmosphere. It can be roughly divided into two categories: controlled - when the spacecraft can determine and influence the time and location of re-entry, and uncontrolled - when the decay is driven by natural forces. To a significant extent, objects demise due to the atmospheric heating but surviving fragments of large-size objects like heavy science satellites can cause a risk to the ground population and on-ground properties.
During 2008-2017 almost 450 large intact objects have re-entered in an uncontrolled way according to \cite{Pardini2019}. This number of objects corresponds to the mass of about 900 metric tons. Therefore, the evolution of an uncontrolled re-entry over a timespan of several years to a couple of hours of the remaining lifetime has to be monitored and predicted.

However, it is challenging to predict the re-entry epoch as there are many factors and uncertainties affecting the accuracy of predictions. The main difficulties are related to inaccurate tracking of objects with a long period between passes, uncertain attitude evolution, calculation of mass, area and drag coefficient of an object, computation of thermosphere density, and future solar and geomagnetic activity \citep{Pardini2013}. 
The thermospheric density is calculated via atmosphere models \citep{Lemmens2016, Virgili2017ISO, Virgili2019} that require as an input the solar extreme ultraviolet radiation and solar wind parameters, such as velocity, density and magnetic field. The rate at which these objects re-enter is not constant as atmospheric decay is the main factor that leads to the re-entry of objects with low-Earth orbits. The drag acting on objects strongly depends on the density of the thermosphere, which changes in response to the UV irradiance and the geomagnetic activity. At an altitude of 400~km, the density increases by an order of magnitude from solar minimum to solar maximum \citep{Emmert2010}. These changes cause large variations in atmospheric drag and therefore determine the frequency of orbiting objects to re-enter the atmosphere.  

\begin{figure}
	\epsscale{0.7}
	\plotone{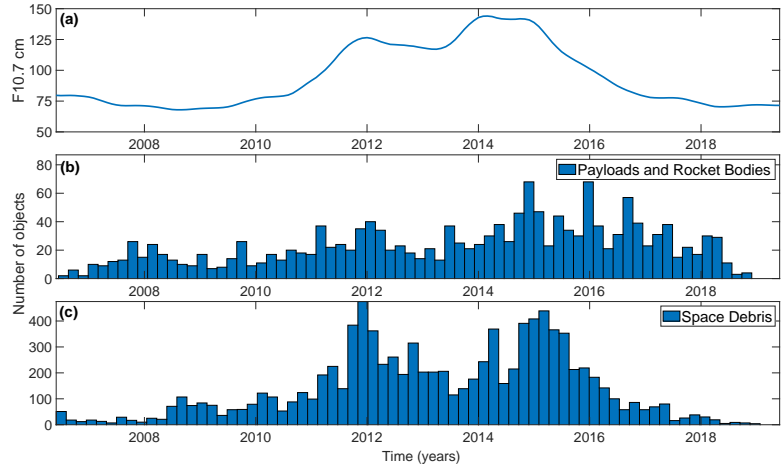}
	\caption{Frequency of re-entered objects as a function of time for solar cycle 24. (a) Smoothed F10.7~cm. (b) Number of objects re-entered for payloads and rocket bodies and (c) space debris.}
	\label{fig9}
\end{figure}

Figure~\ref{fig9} shows the dependence of re-entry frequency on the phase of solar cycle 24 (a) for payloads and rocket bodies (b), and space debris (c). As it can be seen from Figure~\ref{fig9}, the number of re-entered objects for both groups is closely related to the level of solar activity - the majority of objects returned in the phase of the cycle maximum. Consequently, from the view of solar activity predictions, for most of the objects, the accuracy of the re-entry predictions strongly depends on the ability to forecast solar activity in the maximum phase. As shown in Figure~\ref{fig9}(c), the re-entry time of space debris closely follows the cycle evolution, reacting immediately to the changes in the solar activity, whereas payloads and rocket bodies (Figure~\ref{fig9}(b)) show also a large number of re-entries in the area of the second maximum and the declining phase of the cycle, which may be related to the time delay of the re-entry for large objects.

\begin{figure}
	\epsscale{0.8}
	\plotone{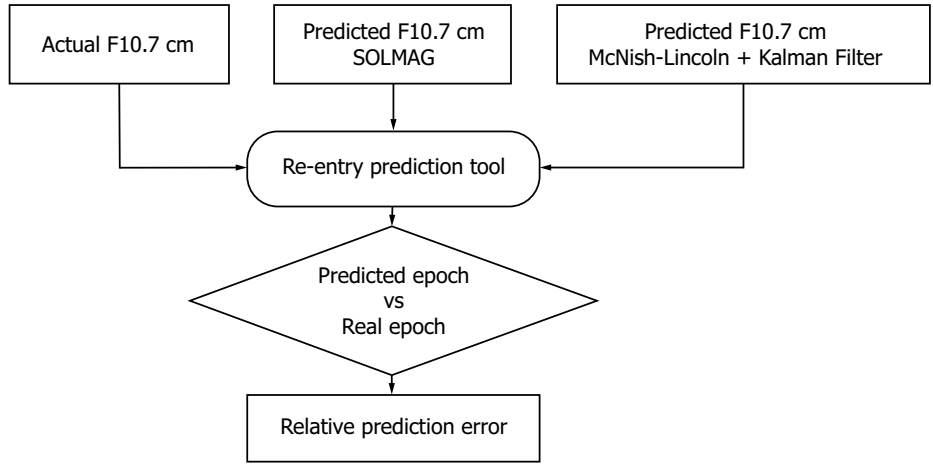}
	\caption{The scheme of the test procedure for re-running the past ESA re-entry campaigns.}
	\label{fig10}
\end{figure}

Figure~\ref{fig10} illustrates the procedure of re-running past ESA re-entry campaigns. 
Actual measurements of F10.7 index ad well as F10.7 predictions by M\&K+KF and SOLMAG can be used as an input to the ESA re-entry prediction tool. As an output, we get the predicted epoch of re-entry and compare it to the actual re-entry time. 

\begin{figure}
	\epsscale{0.5}
	\plotone{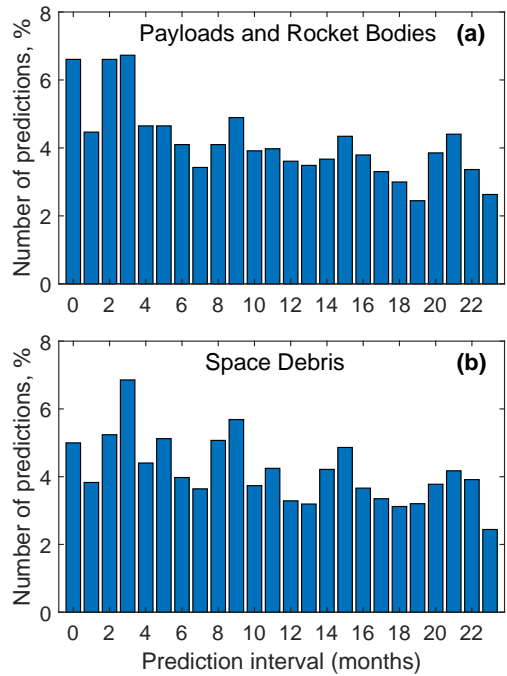}
	\caption{The number of re-entry predictions (in percent) as a function of lead times. (a) Payloads and rocket bodies. (b) Space debris.}
	\label{fig11}
\end{figure}

Over the test period 2006-2019, in total 1635 predictions were made for 602 payloads and rocket bodies, and 9584 predictions for 2344 objects of space debris. The forecast of the re-entry epoch is performed repeatedly for the same object for different lead times. Figure~\ref{fig11} shows how many forecasts are made (in percent) depending on the prediction interval for payloads and rocket bodies (a) and space debris (b). Zero on the X-axis means that the re-entry prediction is made in the same month as an object had re-entered. Although the re-entry prediction interval is less than one month, still one needs to produce a 6-month lead prediction of the smoothed radio flux using the McNish-Lincoln method or to reconstruct the smoothed radio flux value at the current time with the Kalman filter. As there are no re-entry predictions with a 24-month lead time, the last bar in panels~(a)~and~ (b) corresponds to a 23-month re-entry forecast. 

\begin{figure}
	\plotone{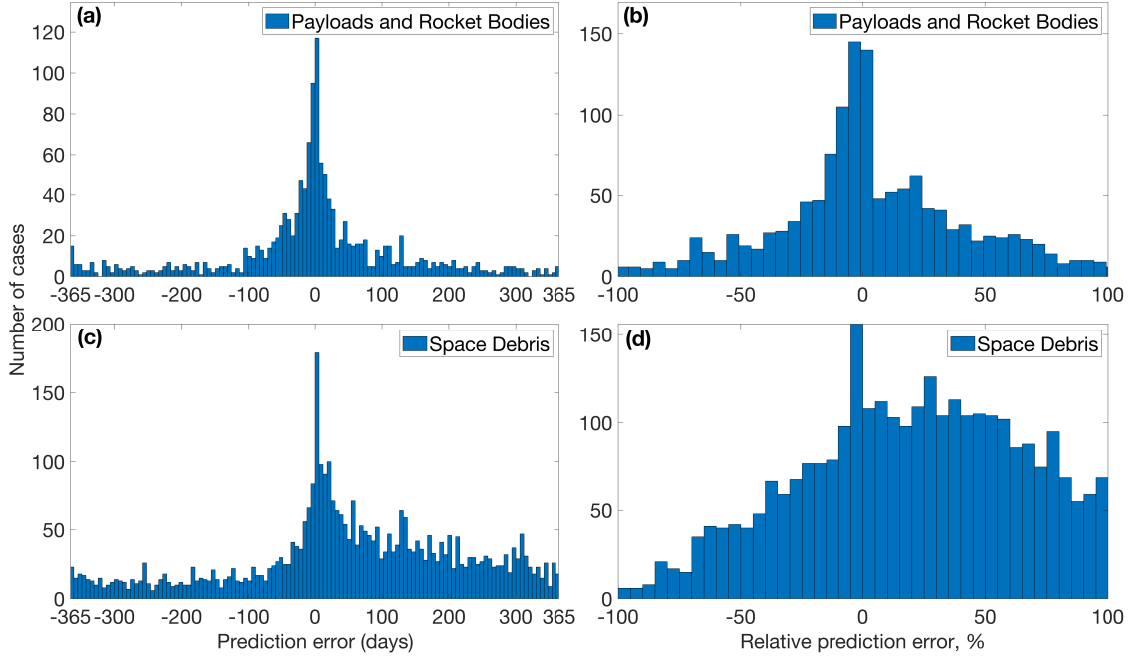}
	\caption{Distribution of re-entry prediction errors for the case of using the actual known F10.7 index as 
		input to the re-entry prediction tool. (a)~and~(b) Payloads and rocket bodies. (c)~and~(b) Space debris.
		The y-axis gives the number of the cases in every category of re-entering objects with a particular prediction error.
		The left panels~(a)~and~(c) show the absolute error in days. The right panels~(b)~and~(d) give the relative prediction error, which is normalized by the lead time (Equation~(\ref{eq:relative_error_reentry})). Positive (negative) errors show cases of the re-entry observed earlier (later) than its prediction.}
	\label{fig12}
\end{figure}

Figure~\ref{fig12} shows the distribution of re-entry prediction errors for the case when the re-entry prediction tool uses the actual known F10.7 as input. The top panels~(a)~and~(b) represent the error distribution for payloads and rocket bodies, and the bottom panels~(c)~and~(d) for objects of space debris. The y-axis indicates the number of the cases in every category of re-entering objects with a particular prediction error. When the error is positive, the predicted re-entry time is later than the actual re-entry, while negative errors show the cases, where the predicted re-entry is earlier than the actual re-entry.  
The left panels~(a)~and~(c) show the absolute error in days. The right panels~(b)~and~(d) give the relative prediction error, which is normalized by the lead time of the predictions and determined in the following way
\begin{equation}\label{eq:relative_error_reentry}
E_{p} = \frac{T_{predicted \, re-entry}-T_{re-entry}}{T_{prediction}-T_{re-entry}}\cdot100\%
\end{equation}
Here, the numerator shows the difference between the epoch of predicted re-entry $T_{predicted \, re-entry}$ and actual re-entry $T_{re-entry}$. The denominator corrects the error by taking into account how far in advance the prediction is made. The larger the interval between the date of the prediction $T_{prediction}$ and actual re-entry $T_{re-entry}$, the less the relative error $E_{p}$. 

As can be seen from the histograms (Figure~\ref{fig12}), the re-entry prediction tool using the actual known F10.7 still provides a large distribution of re-entry prediction errors. Table~\ref{table2} shows the percentage of re-entry predictions with absolute errors from 0 to 100, 100 to 200, 200 to 365, and beyond 365 days (the row ``Absolute error (days)''). Additionally, we show the percentage of re-entry predictions with relative errors from 0 to 33\%, 33 to 66\%, 66 to 100\%, and beyond 100\% (the row ``Relative prediction error, \%''). Most of the re-entry predictions (58\%) in the category of payloads and rocket bodies have an error in the range from 0 to 100 days. However, still, 16\% of cases have an error of re-entry prediction of more than 365 days, and 17\% of cases show a prediction error larger than the lead time. In the category of space debris, the majority of re-entry predictions (61\%) have an error larger than 365 days, and 70\% of cases with the prediction error larger than the lead time. Moreover, for this category, we observe a systematic bias in the re-entry predictions, as in 87\% of the cases, the re-entry is observed earlier than its forecast. It might be related to the wrong assumptions of masses, cross-sections, or other relevant parameters in the drag model that are not well captured for the small debris objects. 
\begin{table*}
 \centering
 \caption{Distribution of re-entry prediction errors. The row ``Absolute error (days)'' gives the percentage of re-entry prediction with with an absolute error in three different intervals from  0 to 365 days, and beyond. The row ``Relative prediction error, \%'' shows the percentage of re-entry predictions with the relative error in three different intervals from 0 to 100\%, and beyond.}
 \label{table2}
 \begin{tabular}{lcccc}
    \hline
	\textbf{Absolute error (days)}          & \textbf{[0;100]}  & \textbf{[100;200]}  & \textbf{[200;365]}  & \boldmath{$>365$} \\ \hline
	Payloads and rocket bodies              & 58                & 13                  & 13                  & 16     \\ \hline
	Space debris                            & 18                & 10                  & 11                  & 61     \\ \hline		
	\textbf{Relative prediction error (\%)} & \textbf{[0;33]}   & \textbf{[33;66]}    & \textbf{[66;100]}   & \boldmath{$>100$} \\ \hline
	Payloads and rocket bodies              & 55                & 19                  & 9                   & 17     \\ \hline
	Space debris                            & 14                & 10                  & 6                   & 70     \\ \hline		
 \end{tabular}
\end{table*}

\begin{figure}
	\plotone{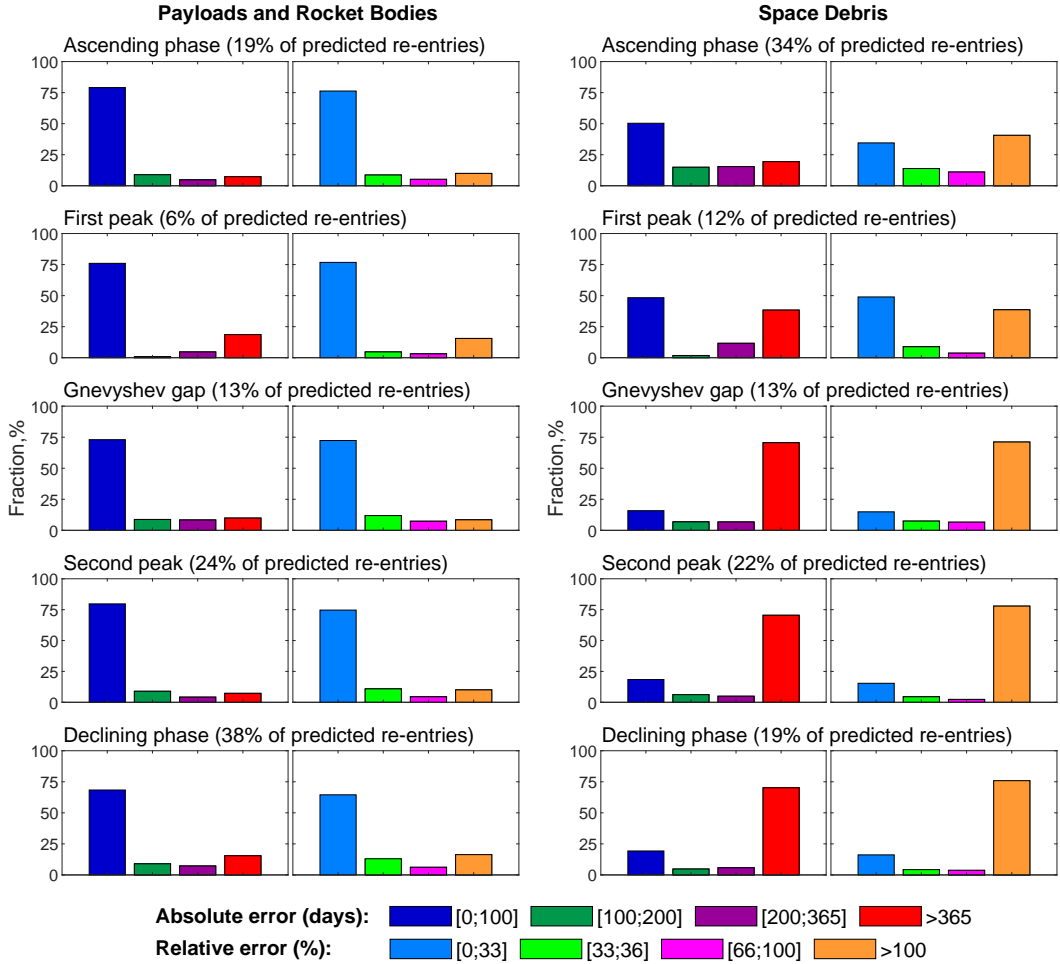}
	\caption{Distribution of re-entry prediction errors for different phases of the solar cycle 24. 
			The cycle is divided into 5 phases: the ascending phase, a segment around the first peak, the Gnevyshev gap, 
			a segment around the second peak, and the declining phase. In brackets, we show the number of predicted re-entries 
			made for a given cycle phase. We show the absolute error (in days) and relative error (in \%) for 
			payloads and rocket bodies (left panels) and for space debris (right panels).
   	        The y-axis gives the percentage of re-entry predictions for each cycle phase with a particular prediction error 
   	        within the ranges indicated in Table~\ref{table2}.}
	\label{fig13}
\end{figure}

Figure~\ref{fig13} shows the re-entry prediction errors for different stages of the solar cycle 24. 
We tentatively divide the cycle into 5 phases: the ascending phase, a segment around the first peak, the Gnevyshev gap, 
a segment around the second peak, and the declining phase. In brackets, we show the number of predicted re-entries 
made for a given cycle phase. The left panels show the absolute error (in days) and relative error (in \%) for 
the payloads and rocket bodies. The right panels give the same, but for the space debris.
The y-axis gives the percentage of re-entry predictions for each cycle phase with a particular prediction error 
within the ranges indicated in Table~\ref{table2}. Figure~\ref{fig13} confirms the results from Figure~\ref{fig12} and 
Table~\ref{table2}, and also shows that the re-entry forecast in the category of space debris has
an error larger than 365 days or exceeding the lead time mainly during Gnevyshev gap, the second peak, and the declining phase 
of the solar cycle. In summary, even perfect knowledge of solar activity does not provide accurate re-entry predictions, and therefore other components of the atmospheric models also need to be improved.

\begin{figure}
	\plotone{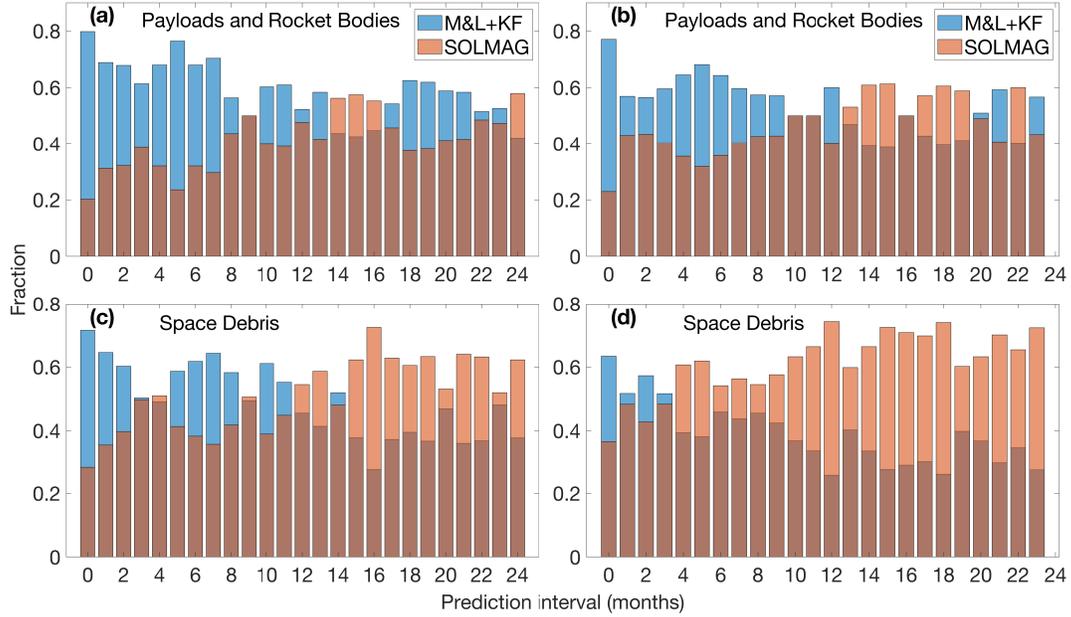}
	\caption{Comparison of the re-entry forecast obtained with the re-entry prediction tool, which uses the F10.7 index predicted by the M\&L+KF and the SOLMAG method. The blue bar shows the fraction of all the forecasts made for each prediction interval where M\&L+KF provides better re-entry predictions. Brown color shows as background the fraction of the non-advantage method, which can be either M\&L+KF or SOLMAG, depending on the forecast lead time. The left panels~(a)~and~(c) show the fraction of cases for the modeled re-entry epoch, while the right ones (b) and (d) for the actual re-entry time. The top panels~(a)~and~(b) give the results for the payloads and rocket bodies, and the bottom panels~(c)~and~(d) for space debris.}
 	\label{fig14}
\end{figure}

To study the influence of the solar activity predictions on the quality of re-entry forecast and to limit the other sources of uncertainty, we use ESA's re-entry tool to model the re-entry epoch with the actual F10.7 index. We further refer to it as ``modeled re-entry epoch''. We then re-run the re-entry prediction tool but use the predicted F10.7 index as input. Then for each prediction interval, we calculate in how many cases the McNish-Lincoln + Kalman filter provides better re-entry predictions than the SOLMAG method as compared to the modeled re-enry epoch, and vice versa. The fraction of all the forecasts made for each prediction interval showing an advantage of the McNish-Lincoln + Kalman filter (M\&L+KF) is presented in Figure~\ref{fig14} by blue color, whereas orange color shows an advantage of the SOLMAG method. Brown color shows as background the fraction of the non-advantage method, which can be either M\&L+KF or SOLMAG, depending on the forecast lead time. The left panels~(a)~and~(c) show the fraction of cases for the modeled re-entry epoch, which allows us to estimate the impact of the F10.7 forecast on the re-entry predictions. For comparison, we also provide the fraction of cases for the actual re-entry time on the right panels~(b)~and~(d), where the predictions are also affected by the unknown deficiencies of the atmospheric models and other input parameters used in the re-entry prediction tool. The top panels~(a)~and~(b) show the results for payloads and rocket bodies, while the bottom panels~(c)~and~(d) for space debris. Note that the last bar in the left panels~(a)~and~(b) corresponds to a 24-month re-entry forecast, as we have the re-entry predictions also with a 24-month lead time for the modeled re-entry, in contrast to the actual re-entry time. 

As can be seen from Figure~\ref{fig14}(a), the M\&L+KF predictions of the F10.7 index used as an input to the re-entry prediction tool, provide a larger fraction of more accurately predicted epochs of re-entry in the category of payloads and rocket bodies. The reason for this is that the majority of predicted re-entries (62\%) in the category of payloads and rocket bodies are observed in the segment around the second cycle peak (24\%) and the declining phase (38\%). This is the area where the M\&L+KF in general outperforms the SOLMAG F10.7 predictions (cf. the heat map in Figure~\ref{fig8}). Some cases that show a larger number of more accurate re-entry predictions with the SOLMAG are related to the interval of Gnevyshev gap (13\% of predicted re-entries), and also partially to the ascending phase and end of cycle 23 (19\% of predicted re-entries), and the first cycle peak (6\% of predicted re-entries).  

For the category of space debris, we see an advantage of the M\&L+KF mainly for the prediction interval 1-11 months ahead, while SOLMAG is mostly better for larger prediction intervals 12-24 months ahead (Figure~\ref{fig14}(c)). The reason for this is that the majority of the predicted re-entries (59\%) are observed in the following three parts of the cycle: ascending phase, the area of the first peak, and the Gnevyshev gap between two maxima. While the percentage of all the predicted re-entries for the Gnevyshev gap between two maxima is the same for both categories, it is increased to 34\% for the ascending phase and to 12\% for the area of the first peak for space debris, which is about twice as large as for the payloads and rocket bodies. At the same time, it is decreased to 22\% for the segment around the second peak and to 19\% for the declining phase. As shown in the heat map (Figure~\ref{fig8}), SOLMAG provides more accurate F10.7 predictions for larger prediction intervals in the late ascending phase and partially around the first cycle peak. This leads to an increase in the number of cases when the re-entry prediction tool, which uses the SOLMAG F10.7 predictions, provides a more accurate forecast of the re-entry epoch. However, as the M\&L+KF shows better performance for these cycle phases in case of a shorter prediction interval (1-11 months ahead), the number of more accurate re-entry predictions with the M\&L+KF for this interval is higher than for the SOLMAG. Note that the impact of both M\&L + KF and SOLMAG radio flux predictions on the accuracy of re-entry forecast depends on the shape of the cycle, and the presence/absence of a significant Gnevyshev gap. The differences in the re-entry predictions for the actual re-entry (right panels (b) and (d) in Figure~\ref{fig14}) is related to other components and uncertainties in the atmospheric models, which are not limited only to the solar activity predictions. Superposition of uncertainties in the atmospheric models and the solar radio flux predictions may lead to improvements of re-entry forecast with the SOLMAG method (Figure~\ref{fig14}d). The further improvement of the re-entry forecast should include both refinements of the atmospheric models and solar activity predictions.

\begin{figure}
	\plotone{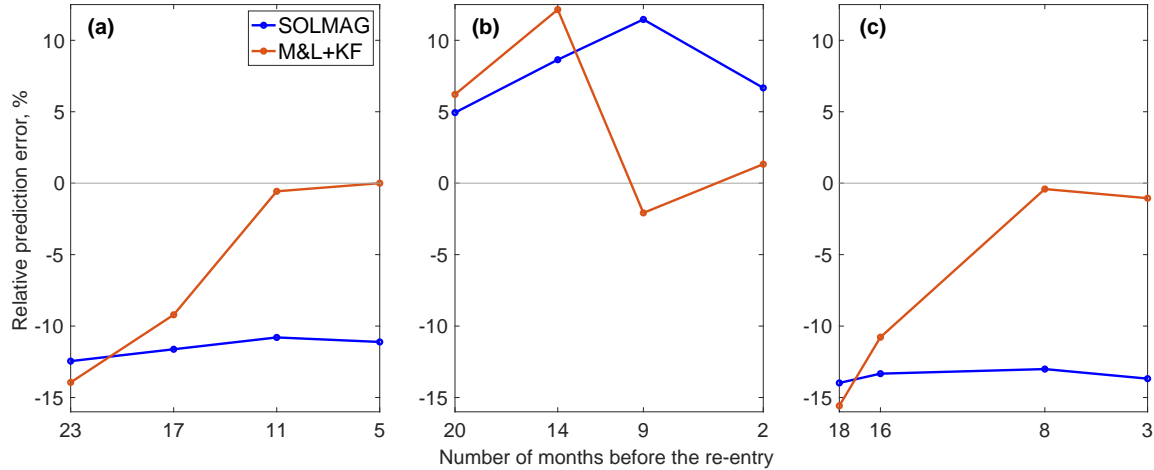}
	\caption{Example of re-entry prediction for three rocket bodies. (a) The second stage of a Delta~3 rocket with a mass of 2476~$kg$ and a cross-section area of 34~$m^{2}$. (b) Centaur, the second stage of an Atlas~1 rocket with a mass of 1840~$kg$ and a cross-section area of 28~$m^{2}$. (c) The third stage of a N-I rocket with a mass of 198~$kg$ and a cross-section area of 2~$m^{2}$. The y-axis shows the relative prediction error in percent derived with Equation~(\ref{eq:relative_error_reentry}). The blue line shows the error of the re-entry predictions with the SOLMAG, while the red line with the M\&L+KF. 
	The positive error shows the cases of the re-entry observed earlier than its prediction. The negative error gives the predictions, which are earlier than the re-entry. The x-axis indicates how far in advance the prediction is made.}
	\label{fig15}
\end{figure}

Figure~\ref{fig15} shows an example of re-entry prediction for the second stage of a Delta~3 rocket (a), Centaur, the second stage of an Atlas~1 rocket (b), and the third stage of a N-I rocket (c). The blue line indicates the relative error of the re-entry predictions (Equation~(\ref{eq:relative_error_reentry})), which use the SOLMAG F10.7 predictions as input to the re-entry prediction tool. The red line shows the same for the M\&L+KF F10.7 predictions as input.

As can be seen from Figure~\ref{fig15}(a), the accuracy of re-entry prediction done 23 months before the re-entry of the second stage of the Delta~3 rocket is comparable for both SOLMAG (blue line) and M\&L+KF (red line). However, closer to the re-entry, the prediction error related to the M\&L+KF strongly decreases (almost to zero), while it is still big for SOLMAG. The modeled re-entry epoch is 29 December 2008. The 5-month lead forecast predicts that the object will re-enter on 10 December 2008 for SOLMAG and on 18 December 2008 for M\&L+KF, which leads to an increase of re-entry prediction accuracy by 18 days. The relative prediction error with M\&L+KF dropped almost to zero. The error of re-entry prediction for the Centaur (Figure~\ref{fig15}(b)) increases with the decrease of prediction advance time from 20 to 14 months, and SOLMAG shows better performance compared to M\&L+KF at these intervals. However, closer to the re-entry, for instance the error of the 9-month lead forecast for M\&L+KF significantly decreases, while that for SOLMAG still increases. The modeled re-entry epoch for this object is 20 October 2014. According to the 9-month lead forecast, the object will re-enter on 22 November 2014 for SOLMAG, which gives more than a 1-month delayed prediction (the relative prediction error is 11.5\%). At the same time, M\&L+KF predicts the re-entry to happen on 14 October 2014, which provides an increase of re-entry prediction accuracy by 27 days. The relative prediction error with M\&L+KF decreases to -2.1\%, which is about 5 times smaller than that for SOLMAG. The dynamics of the re-entry prediction errors for the third stage of the N-I rocket (Figure~\ref{fig15}(c)) is similar to that of the second stage of the Delta~3 rocket (Figure~\ref{fig15}(a)). For instance, the 8-month lead forecast shows that the object will re-enter on 5 February 2009 for SOLMAG and on 8 March 2009 for M\&L+KF. The modeled re-entry epoch is 9 March 2009. Thus, the use of the M\&L+KF provides the re-entry prediction 31 days more accurate than SOLMAG. The relative prediction error with M\&L+KF decreases to -0.4\%, which is 32 times smaller than that for SOLMAG. These cases reflect the general trend, that for longer lead times SOLMAG provides better predictions, whereas for shorter lead times, i.e. when the re-entry comes closer, M\&L+KF shows better prediction results (cf. Figure~\ref{fig14}).

\section{Discussion and Conclusions}
The findings and outcomes of this study can be divided into two main groups. First, we developed a prediction techniques (M\&L+KF) for the F10.7 and F30 indices with lead times of 1-24 months ahead and compared the performance of our approach with the current state-of-the-art SOLMAG prediction method used by ESA. Second, we tested the impact of the developed techniques on the application case of re-entry predictions through re-running past ESA re-entry campaigns for payloads, rocket bodies, and space debris that re-entered over the years 2006-2016, covering in total 602 payloads and rocket bodies, and 2344 objects of space debris. 

The developed M\&L+KF prediction technique of the solar radio flux consists of three main steps. First, to process the short-term fluctuating component, which should be removed to detect the trend, we applied a 13-month optimized running mean technique \citep{Podladchikova2017}. It results in a better forecasting accuracy compared to the traditional 13-month running mean with a reduction of the prediction errors by 13\% for F10.7 and 19\% for F30. Second, we used the McNish-Lincoln method to provide initial predictions of the radio flux indices. However, the 13-month running mean of the radio flux, which is used as input to the McNish-Lincoln method, has a 6-month delay with respect to the current time. The radio flux activity over the last 6-month period until the current time is considered to be unknown because of the noise component intrinsic to the available monthly mean data. To remove this drawback, we developed an adaptive Kalman filter with noise statistics identification, which assimilates the monthly mean radio flux data over the last 6-month period and provides an estimation of the smoothed radio flux at the current time. The reconstructed radio flux at the current time becomes a new starting point for the predictions, which allows the McNish-Lincoln (M\&L) method to perform without a 6-month delay. We showed that combining the McNish-Lincoln method with the Kalman filter (M\&L+KF) improves the prediction accuracy by 36\% for F10.7 and 39\% for F30 compared to the initial M\&L predictions.  

It is important to mention, that the proposed Kalman filter technique assimilating the monthly mean radio flux data over the last 6-month period represents an universal approach, as its application is independent of the original prediction model. This approach can be applied not only to the McNish-Lincoln predictions but also to the radio flux predictions provided by any other forecasting method. It might be of practical interest in the future to develop and test the proposed approach for the Standard and Combined methods, initially designed to predict sunspot numbers. A comparative study of the performance of these methods developed for solar radio flux prediction would be of practical importance to test the forecasting capacities for cycles with different shapes and behaviors. 
Potentially, the proposed approach can be applied to any other proxy of solar activity, however, different observables require specific solutions and tuning.

In the current study, the developed M\&L+KF method gives a RMSE of predictions in the range of of 5-27~sfu for F10.7 and 3-16~sfu for F30. This result statistically outperforms the current state-of-the-art SOLMAG  prediction method employed by ESA by 15.5-66.5\% and can be recommended for the implementation at ESA.

Additionally, we performed a systematic evaluation of the re-entry predictions through re-running past ESA re-entry campaigns for payloads, rocket bodies, and objects of space debris that re-entered over the last solar cycle.
We showed that even using the known F10.7 as input to the re-entry prediction tool, 16\% of the cases for payloads and rocket bodies, and 61\% of the cases for space debris, have an absolute error of re-entry prediction of more than 365 days. The percentage of cases where the relative prediction error is larger than the lead time is 17\% for the payloads and rocket bodies, and 70\% for the space debris. Moreover, in the category of space debris, we observe a systematic bias in the re-entry predictions, as in 87\% of the cases the re-entry is observed earlier than its forecast. These results can help to improve the handling of atmospheric model parameters (assumptions of object masses, cross-sections, etc.) in the future and to improve the re-entry prediction tool.

We also tested the influence of the solar radio flux predictions on the accuracy of the re-entry forecast, when the predicted F10.7 was used as input to the re-entry prediction tool. We showed that F10.7 predicted by the proposed M\&K+KF method provides a larger fraction of more accurate re-entry forecast in the category of payloads and rocket bodies compared to the SOLMAG method. For the category of space debris, M\&L+KF shows an advantage in the re-entry forecast for the prediction interval 1-11 months ahead, while SOLMAG is mostly better for larger prediction intervals 12-24 months ahead. This is partially related to the accuracy of the F10.7 predictions by M\&L+KF and SOLMAG in cycle 24, which has a prominent Gnevyshev gap. 

Further improvements of re-entry forecasts need to include both, refinements of the atmospheric models (assumptions of object masses, cross-sections, etc.) as well as of solar activity predictions. From our study, it is clear that even with ``perfect input'' on the solar activity (i.e. the measured instead of the predicted F10.7 index), there are still substantial uncertainties in the re-entry predictions, which hints at the other drag model components. As for improvements of the medium-term solar activity / radio flux predictions, it is highly important to have a better forecast not only of the peak but also the shape of a solar cycle. In that respect, a better understanding and prediction of the Gnevyshev gaps is needed, as those determine the activity amplitude/phase during the period of high solar activity, where the drag affecting on space objects is the strongest.

Note as was by \citet{Dudok2014}, the F30 index can be a better proxy for modeling the thermosphere-ionosphere system than the currently used 10.7 index. At the same time, as was shown in this study, the predication accuracy of F30 itself is higher than that of F10.7 (Figures~\ref{fig4}~and~\ref{fig6}). It is related to the lower variability of F30 compared to F10.7 with the variance of monthly mean F30 data being almost 3 times smaller than that of F10.7. Thus, the incorporation of the F30 index into the re-entry prediction tool is expected to improve the forecast of the re-entry epoch.

\acknowledgments
The authors acknowledge acknowledge the Nobeyama Radio Observatory (Nobeyama, Japan) and the Dominion Radio Astrophysical observatory (Penticton, Canada) for providing F10.7 and F30 indices, the team of ESA's Space Debris Office for the SOLMAG predictions and data of the past ESA re-entry campaigns, and WDC-SILSO at the Royal Observatory of Belgium (ROB) for the sunspot number data sets. We thank the referee for valuable comments on this study. A.M.V. acknowledges the support of the Austrian Science Fund (FWF): P27292-N20 and I4555. 

\appendix
\section{Error derivation for McNish-Lincoln method}\label{ML_error}
To derive the errors of the McNish-Lincoln predictions we do the following estimations: 

\subsection{The variance of smoothed radio flux data} 
Let $F_{m}^{c}$ denote the values of the smoothed radio flux data at any month $m$ of a cycle $c$. As the mean cycle in the McNish-Lincoln method is constructed starting from cycle 8, $c$ changes from 8 to $N_{c}$, where $N_{c}$ is the number of the last available cycle. We assume that the sequence  $F_{m}^{c},\ (c=8,9, \ldots, N_{c})$ is uncorrelated. Then the mean value, $\bar{F}_{m}$, of smoothed radio flux data at month $m$ over the considered cycles $c$ is estimated as 
\begin{equation}\label{eq_A1}
\bar{F}_{m}=\frac{1}{N_{c}} \sum_{c=8}^{N_{c}}F_{m}^{c}  
\end{equation}
The variance of $F_{m}^{c}$ is then represented by
\begin{equation}\label{eq_A2}
\sigma_{m}^{2}=\frac{1}{N_{c}-1}\sum_{c=8}^{N_{c}}E\left[\left(F_{m}^{c}-\bar{F}_{m}\right)^{2}\right] =\frac{1}{N_{c}-1}\sum_{c=8}^{N_{c}}\left(E\left[\left(F_{m}^{c}\right)^{2}\right]-E\left[2\bar{F}_{m}F_{m}^{c}\right]+E\left
[\bar{F}_{m}^{2}\right]\right) 
\end{equation}
Here, the symbol $E$ denotes the mathematical expectation.
Let us consider every term of Equation~(\ref{eq_A2}) separately. \\
\textit{Term 1:}
\begin{equation}\label{eq_A3}
\frac{1}{N_{c}-1}\sum_{c=8}^{N_{c}}E\left[\left(F_{m}^{c}\right)^{2}\right]=E\left[\frac{1}{N_{c}-1}\sum_{c=8}^{N_{c}}\left(F_{m}^{c} \right)^{2}\right] = \frac{1}{N_{c}-1}\sum_{c=8}^{N_{c}}\left(F_{m}^{c}\right)^{2}
\end{equation}
Similarly, \textit{term 2} can be estimated as \\
\begin{equation}\label{eq_A4}
\frac{1}{N_{c}-1}\sum_{c=8}^{N_{c}}E\left[2\bar{F}_{m}F_{m}^{c}\right] =\frac{1}{N_{c}-1}2\bar{F}_{m}\sum_{c=8}^{N_{c}}F_{m}^{c} 
\end{equation}
Taking into account Equation~(\ref{eq_A1})
\begin{equation}\label{eq_A5}
\sum_{c=8}^{N_{c}}F_{m}^{c}=N_{c}\bar{S}_{m}
\end{equation}
Then, term 2 can be rewritten as
\begin{equation}\label{eq_A6}
\frac{1}{N_{c}-1}\sum_{c=8}^{N_{c}}E\left[2\bar{F}_{m}F_{m}^{c}\right]=\frac{1}{N_{c}-1}2N_{c}\bar{F}_{m}^{2}
\end{equation}
\textit{Term 3:}\\
\begin{equation}\label{eq_A7}
\frac{1}{N_{c}-1}\sum_{c=8}^{N_{c}}E\left[\bar{F}_{m}^{2}\right] =\frac{1}{N_{c}-1}\bar{F}_{m}^{2}\sum_{c=8}^{N_{c}}1=\frac{1}{N_{c}-1}N_{c}\bar{F}_{m}^{2} 
\end{equation}
Thus, Equation~(\ref{eq_A2}) representing the variance $\sigma_{m}^2$ can be rewritten as
\begin{equation}\label{eq_A8}
\sigma_{m}^{2}=\frac{1}{N_{c}-1}\left[\sum_{c=8}^{N_{c}}\left(F_{m}^{c}\right)^{2}-2N_{c}\bar{F}_{m}^{2}+N_{c}\bar{F}_{m}^{2} \right] 
\end{equation}
or
\begin{equation}\label{eq_A9}
\sigma_{m}^{2}=\frac{\sum_{c=8}^{N_{c}}\left(F_{m}^{c}\right)^{2}-N_{c}\bar{F}_{m}^{2}}{N_{c}-1}
\end{equation}
Similarly to Equation~(\ref{eq_A9}), the variance $\sigma_{i}^2$ is given by
\begin{equation}\label{eq_A10}
\sigma_{i}^{2}=\frac{\sum_{c=8}^{N_{c}}\left(F_{i}^{c}\right)^{2}-N_{c}\bar{F}_{i}^{2}}{N_{c}-1}
\end{equation}
Here, $F_{i}^{c}$ denote the values of smoothed radio flux data at any month $i$ of a cycle $c,\ (c=8,9, \ldots, N_{c})$, 
and $\bar{F}_{i}$ represents the mean value of smoothed radio flux data at month $i$ over the considered cycles $c$. 

\subsection{Construction of linear regression}
Let us introduce the following residuals denoting the difference between the smoothed radio flux value and the value of the mean cycle at months $m$ and $i$
\begin{equation}\label{eq_A11}
	\Delta_{m}^{c}=F_{m}^{c}-\bar{F}_{m}, \qquad (c = 8,9,... , N_{c})
\end{equation}
\begin{equation}\label{eq_A12}
	\Delta_{i}^{c}=F_{i}^{c}-\bar{F}_{i}, \qquad (c = 8,9,... , N_{c})
\end{equation}
Then we form the following linear regression equation between $\Delta_{m}^{c}$ and $\Delta_{i}^{c}$
\begin{equation}\label{eq_A13}
\Delta_{m}^{c}=a_{mi}+k_{mi} \Delta_{i}^{c}+\varepsilon_{mi}^{c}
\end{equation}
Here, $\varepsilon_{mi}^{c}$ is the uncorrelated noise representing the model error.
Let us introduce the vector
\begin{equation}\label{eq_A14}
Y=  \begin{vmatrix}
     \Delta_{m}^{8}  & \Delta _{m}^{9}  & \cdots   & \Delta _{m}^{N_{c}}\\
     \end{vmatrix}
   ^{T},
\end{equation}
the vector of estimated parameters
\begin{equation}\label{eq_A15}
X_{mi}= \begin{vmatrix}
        a_{mi}\\
        k_{mi}
        \end{vmatrix},
\end{equation}
the matrix
\begin{equation}\label{eq_A16}
H= \begin{vmatrix}
   1                & \ldots   & 1\\
   \Delta _{i}^{8}  & \ldots   &  \Delta _{i}^{N_{c}}\\
\end{vmatrix}^{T},
\end{equation}
and vector 
\begin{equation}\label{eq_A17}
\varepsilon =  \begin{vmatrix}
               \varepsilon _{mi}^{8}  & \varepsilon_{mi}^{9}  & \cdots   & \varepsilon _{m}^{N_{c}}\\
               \end{vmatrix}
^{T}.
\end{equation}
Then the regression Equation \ref{eq_A13} can be rewritten as 
\begin{equation}\label{eq_A18}
   Y=HX_{mi}+ \varepsilon 
\end{equation}
Here, $\sigma_{\varepsilon}^{2}$  is the  variance of noise $\varepsilon$ representing the scatter around the regression line.
 
We determine the estimate of vector  $X_{mi}$  with unknown parameters $a_{mi}$ and $k_{mi}$ on the basis of the least-square method
\begin{equation}\label{eq_A19}
\hat{X}_{mi}= \begin{vmatrix}
              \hat{a}_{mi}\\
              \hat{k}_{mi}\\
              \end{vmatrix}
=\left(H^{T}H\right)^{-1}H^{T}Y 
\end{equation}
Then Equation~(\ref{eq_A19}) can be rewritten as 
\begin{equation}\label{eq_A20}
\hat{X}_{mi} =  \begin{vmatrix}
    		    \hat{a}_{mi}\\
		        \hat{k}_{mi}\\
	            \end{vmatrix}
             = \begin{vmatrix}
		        0\\
		       \frac{\sum_{c=8}^{N_{c}}\Delta_{m}^{c}\Delta_{i}^{c}}{\sum_{c=8}^{N_{c}}\left(\Delta_{i}^{c}\right)^{2}}\\
 	           \end{vmatrix}
\end{equation}
Thus, the unknown parameters of regression can be determined as
\begin{equation}\label{eq_A21}
    \hat{a}_{mi}=0 
\end{equation}
\begin{equation}\label{eq_A22}
\hat{k}_{mi}=\frac{\sum_{c=8}^{N_{c}}\Delta_{m}^{c}\Delta_{i}^{c}}{\sum_{c=8}^{N_{c}}\left(\Delta_{i}^{c}\right)^{2}}
\end{equation}

\subsection{Variances of estimation errors of coefficients $\hat{a}_{mi}$ and $\hat{k}_{mi}$}
Let us introduce the matrix 
\begin{equation}\label{eq_A23}
D=\left(H^{T}H\right)^{-1}H^{T}=\begin{vmatrix}
  \frac{1}{N_{c}}  & \ldots   & \frac{1}{N_{c}}\\
  \frac{\Delta_{i}^{8}}{\sum_{c=8}^{N_{c}}\left(\Delta_{i}^{c}\right)^{2}}  & \ldots   & \frac{\Delta_{i}^{N_{c}}}{\sum_{c=8}^{N_{c}}\left(\Delta_{i}^{c}\right)^{2}}\\
\end{vmatrix}
\end{equation}
Then we can write
\begin{equation}\label{eq_A24}
\hat{X}_{mi}=  \begin{vmatrix}
               \hat{a}_{mi}\\
               \hat{k}_{mi}\\
               \end{vmatrix}
=DY 
\end{equation}
The covariance matrix of the estimation error of $\hat{X}_{mi}$ is given by
\begin{equation}\label{eq_A25}
cov\left(\hat{X}_{mi}\right)=Dvar\left(Y\right)D^{T}=\sigma_{\varepsilon }^{2}DD^{T}=\sigma_{ \varepsilon }^{2}  
\begin{vmatrix}
\frac{1}{N_{c}}  &  0\\
0                &  \frac{1}{\sum_{c=8}^{N_{c}}\left(\Delta_{i}^{c}\right)^{2}}\\
\end{vmatrix}
\end{equation}
The variance of $\Delta_{i}^{c}$ is given by
\begin{equation}\label{eq_A26}
     \sigma_{i}^{2}=\frac{1}{N_{c}-1}\sum_{c=8}^{N_{c}}\left(\Delta_{i}^{c}\right)^{2} 
\end{equation}
and then
\begin{equation}\label{eq_A27}
 \sum_{c=8}^{N_{c}} \left(\Delta_{i}^{c}\right)^{2}= \sigma_{i}^{2}\left(N_{c}-1\right)
\end{equation}
Thus, taking into account Equation~(\ref{eq_A27}),  Equation~(\ref{eq_A25}) can be rewritten as 
\begin{equation}\label{eq_A28}
cov\left(\hat{X}_{mi}\right)=
\begin{vmatrix}
    \frac{1}{N_{c}}  &  0\\
    0                &  \frac{1}{\sigma_{i}^{2}\left(N_{c}-1\right) }\\
\end{vmatrix}
\end{equation}
Therefore, the variances of estimates $\hat{a}_{mi}$ and $\hat{k}_{mi}$ are presented as 
\begin{equation}\label{eq_A29}
var\left(\hat{a}_{mi}\right) = \sigma_{\varepsilon}^{2}\frac{1}{N_{c}}
\end{equation}
\begin{equation}\label{eq_A30}
var\left(\hat{k}_{mi}\right) = \sigma_{\varepsilon}^{2}\frac{1}{\sigma_{i}^{2}\left(N_{c}-1\right)}
\end{equation}

\subsection{The variance of prediction errors} 
Let us rewrite Equation~(\ref{eq_A13}) in the following way
\begin{equation}\label{eq_A31}
F_{m}^{c}=\bar{F}_{m}+a_{mi}+k_{mi}\left(F_{i}^{c}-\bar{F}_{i}\right) + \varepsilon_{mi}^{c} 
\end{equation}
The forecast to month $m$ from cycle $c$ is then made by  
\begin{equation}\label{eq_A32}
\hat{F}_{m}^{c}=\bar{F}_{m}+\hat{a}_{mi}+\hat{k}_{mi}\left(F_{i}^{c}-\bar{F{i}}\right)
\end{equation}
The prediction error $\delta_{im}$ at month $m$ is then determined by subtracting Equation~(\ref{eq_A32}) from Equation~(\ref{eq_A31})
\begin{equation}\label{eq_A33}
\begin{multlined}
\delta_{im}=F_{m}^{c}-\hat{F}_{m}^{c}=\left(a_{mi}-\hat{a}_{mi}\right)+k_{mi}\left(F_{i}^{c}-\bar{F_{i}}\right)+
\varepsilon_{mi}-\hat{k}_{mi}\left(F_{i}^{c}-\bar{F_{i}} \right) =\\
 =\varepsilon_{mi}+ \left(a_{mi}-\hat{a}_{mi}\right) + \left(k_{mi}-\hat{k}_{mi}\right)\left(F_{i}^{c}-\bar{F_{i}}\right)  
\end{multlined}
\end{equation}
The variance $\hat{\sigma}_{p}^{2}$ of prediction error $\delta_{im}$ is given by
\begin{equation}\label{eq_A34}
\hat{\sigma}_{p}^{2}=var\left(\hat{\delta}_{mi}\right)=\hat{\sigma }_{\varepsilon}^{2}+var\left(\hat{a}_{mi}\right)+
var\left(\hat{k}_{mi}\right)\left(F_{i}^{c}-\bar{F_{i}}\right)^{2} 
\end{equation}
or 
\begin{equation}\label{eq_A35}
\hat{\sigma}_{p}^{2}=\hat{\sigma}_{\varepsilon}^{2}\left(1+\frac{1}{N_{c}}+\frac{\left(F_{i}^{c}-\bar{F_{i}}\right)^{2}}
{\sigma_{i}^{2}\left(N_{c}-1\right)}\right) 
\end{equation}

\subsection{The estimation of standard deviation $\hat{\sigma}_{\varepsilon}$}
As follows from Equation~(\ref{eq_A13})
\begin{equation}\label{eq_A36}
\varepsilon_{mi}^{c}= \Delta_{m}^{c}-k_{mi}\Delta_{i}^{c}-a_{mi}
\end{equation}
The variance estimation of noise  $\varepsilon_{mi}^{c}$ is presented as
\begin{equation}\label{eq_A37}
\hat{\sigma }_{\varepsilon}^{2}=\frac{\sum_{c=8}^{N_{c}}\left(\Delta_{m}^{c}-k_{mi}\Delta_{i}^{c}\right)^{2}}{N_{c}-2}
\end{equation}
Here we use $N_{c}-2$ in the denominator, as we estimate two regression coefficients  $a_{mi}$ and $k_{mi}$. 
As follows from Equation~(\ref{eq_A37})
\begin{equation}\label{eq_A38}
\begin{multlined}
\hat{\sigma }_{\varepsilon }^{2}=\frac{\sum_{c=8}^{N_{c}}\left(\left(\Delta_{m}^{c}\right)^{2}-2k_{mi}\Delta_{m}^{c}
\Delta_{i}^{c}+\left(k_{mi}\Delta_{i}^{c}\right)^{2}\right)}{N_{c}-2}=\\
=\frac{\sum_{c=8}^{N_{c}}\left(\Delta_{m}^{c}\right)^{2}-2k_{mi}\sum_{c=8}^{N_{c}}\Delta_{m}^{c}\Delta_{i}^{c}+k_{mi}^{2}
\sum_{c=8}^{N_{c}}\left(\Delta_{i}^{c}\right)^{2}}{N_{c}-2}
\end{multlined}
\end{equation}
Let us consider all the terms in the nominator of Equation~(\ref{eq_A38}). According to Equation (\ref{eq_A27}), \textit{the first term} can be presented as 
\begin{equation}\label{eq_A39}
\sum_{c=8}^{N_{c}}\left(\Delta_{m}^{c}\right)^{2}= \sigma_{m}^{2}\left(N_{c}-1\right)
\end{equation}
Let us multiply and divide \textit{the second term} on $\sum_{c=8}^{N_{c}}\left(\Delta_{i}^{c}\right)^{2}$.  Then
\begin{equation}\label{eq_A40}
 2k_{mi}\sum_{c=8}^{N_{c}}\Delta_{m}^{c}\Delta_{i}^{c}=2k_{mi}\frac{\sum_{c=8}^{N_{c}}\Delta_{m}^{c}\Delta_{i}^{c}}
 {\sum_{c=8}^{N_{c}}\left(\Delta_{i}^{c}\right)^{2}}\sum_{c=8}^{N_{c}}\left(\Delta_{i}^{c}\right)^{2} 
\end{equation}
Taking into account Equations~(\ref{eq_A22})~and(\ref{eq_A27}), we can rewrite Equation~(\ref{eq_A40}) in the following way
\begin{equation}\label{eq_A41}
 2k_{mi}\sum_{c=8}^{N_{c}}\Delta_{m}^{c}\Delta_{i}^{c}=2k_{mi}^{2}\sigma_{i}^{2}\left(N_{c}-1\right)
\end{equation}
As follows from Equation~(\ref{eq_A41}), the variance $\sigma _{i}^{2}$ is given by 
\begin{equation}\label{eq_A42}
\sigma_{i}^{2}=\frac{2k_{mi}^{2}\left(N_{c}-1\right)}{2k_{mi}\sum_{c=8}^{N_{c}}\Delta_{m}^{c}\Delta_{i}^{c}}=
\frac{k_{mi}\left(N_{c}-1\right)}{\sum_{c=8}^{N_{c}}\Delta_{m}^{c}\Delta_{i}^{c}}=\frac{\frac{\sum _{c=8}^{N_{c}}
\Delta_{m}^{c}\Delta_{i}^{c}}{\sum_{c=8}^{N_{c}}\left(\Delta_{i}^{c}\right)^{2}}\left(N_{c}-1\right)}
{\sum_{c=8}^{N_{c}}\Delta_{m}^{c}\Delta_{i}^{c}}
\end{equation}
Thus, taking into acount Equation~(\ref{eq_A10}) 
\begin{equation}\label{eq_A43}
\sigma_{i}^{2}=\frac{\sum_{c=8}^{N_{c}}\left(F_{i}^{c}\right)^{2}-N_{c}\left(\bar{F}_{i}\right)^{2}}{N_{c}-1}=
\frac{\sum_{c=8}^{N_{c}}\left(\Delta_{i}^{c}\right)^{2}}{N_{c}-1} 
\end{equation}
Using Equaion~(\ref{eq_A27}), we can rewrite \textit{the third term} of Equation~(\ref{eq_A38}) in the following way
\begin{equation}\label{eq_A44}
\left(k_{mi}\right)^{2}\sum_{c=8}^{N_{c}}\left(\Delta_{i}^{c}\right)^{2}=
\left(k_{mi}\right)^{2}\sigma_{i}^{2}\left(N_{c}-1\right)
\end{equation}
Thus, Equation~(\ref{eq_A38}) can be rewritten as 
\begin{equation}\label{eq_A45}
\hat{\sigma }_{\varepsilon}^{2}=\frac{\left(\sigma_{m}^{2}-2k_{mi}^{2}\sigma_{i}^{2}+k_{mi}^{2}\sigma_{i}^{2}\right)
\left(N_{c}-1\right)}{N_{c}-2}=\frac{\left(\sigma_{m}^{2}-k_{mi}^{2}\sigma_{i}^{2}\right)\left(N_{c}-1 \right)}{N_{c}-2}
\end{equation}
The standard deviation $\hat{\sigma}_{\varepsilon}$ is then determined as
\begin{equation}\label{eq_A46}
\hat{\sigma}_{\varepsilon}=\sqrt{\frac{\left(\sigma_{m}^{2}-k_{mi}^{2}\sigma_{i}^{2}\right)\left(N_{c}-1\right)}{N_{c}-2}}
\end{equation}
Taking into account Equation~(\ref{eq_A35}), the standard deviation of the prediction is given by
\begin{equation}\label{eq_A47}
\hat{\sigma}_{p}=\sqrt{\frac{\left(\sigma_{m}^{2}-k_{mi}^{2}\sigma_{i}^{2}\right)\left(N_{c}-1\right)}{N_{c}-2}}
\sqrt{1+\frac{1}{N_{c}}+\frac{\left(F_{i}^{c}-\bar{F_{i}}\right)^{2}}{\sigma_{i}^{2}\left(N_{c}-1\right)}}
\end{equation}
Here, $k_{mi}$ is derived according to Equation~(\ref{eq_A22}).
 
\subsection{Error derivation for McNish-Lincoln+Kalman filter}\label{MLKF_error}
In this section we derive the error of predictions obtained by the combination of the McNish-Lincoln method with the Kalman filter.
The forecast to month $m$ from cycle $c$ we use Equation~(\ref{eq_A32}).
However, here, we use the estimated $\hat{F}_{i}^{c}$ with the Kalman filter instead of the available smoothed radio flux ${F}_{i}^{c}$. We can then present ${F}_{i}^{c}$ as   
\begin{equation}\label{eq_A48}
{F}_{i}^{c}=\hat{F}_{i}^{c}-\zeta
\end{equation}
Here $\zeta$ is the random noise with variance $\sigma_{i,i}^2$ derived from Equation~(\ref{KF_filtration_error}), which
describes the estimation accuracy of smoothed radio flux data ${F}_{i}^{c}$.
Rewriting Equation~(\ref{eq_A33}), we obtain
\begin{equation}\label{eq_A49}
\begin{multlined}
\delta_{im}=F_{m}^{c}-\hat{F}_{m}^{c}=\left(a_{mi}-\hat{a}_{mi}\right)+k_{mi}\left(\hat{F}_{i}^{c}-\zeta-\bar{F_{i}}\right)+
\varepsilon_{mi}-\hat{k}_{mi}\left(\hat{F}_{i}^{c}-\zeta-\bar{F_{i}} \right) =\\
=\varepsilon_{mi}+ \left(a_{mi}-\hat{a}_{mi}\right) + \left(k_{mi}-\hat{k}_{mi}\right)\left( F_{i}^{c}-\bar{F_{i}}\right)-\hat{k}_{mi}\zeta
\end{multlined}
\end{equation}
The variance $\hat{\sigma}_{p}^{2}$ of prediction error $\delta_{im}$ is given by
\begin{equation}\label{eq_A50}
\hat{\sigma}_{p}^{2}=\hat{\sigma}_{\varepsilon}^{2}\left(1+\frac{1}{N_{c}}+\frac{\left(\hat{F}_{i}^{c}-\bar{F_{i}}\right)^{2}}
{\sigma_{i}^{2}\left(N_{c}-1\right)}\right)+k_{mi}^{2}\sigma_{i,i}^2 
\end{equation}
Here, $k_{mi}^{2}$ is estimated using Equation~(\ref{eq_A22}) replacing $\hat{k}_{mi}^{2}$ for convenience. 
 
Then, taking into account Equations~(\ref{eq_A46})~and (\ref{eq_A50}), the standard deviation of the McNish-Lincoln+Kalman filter prediction is given by
\begin{equation}\label{eq_A51}
\hat{\sigma}_{p}=\sqrt{\frac{\left(\sigma_{m}^{2}-k_{mi}^{2}\sigma_{i}^{2}\right)\left(N_{c}-1\right)}{N_{c}-2}
\left(1+\frac{1}{N_{c}}+\frac{\left(\hat{F}_{i}^{c}-\bar{F_{i}}\right)^{2}}{\sigma_{i}^{2}\left(N_{c}-1\right)}\right)+k_{mi}^{2}\sigma_{i,i}^2}
\end{equation}
Note, that Equation~(\ref{eq_A51}) contains an additional term $k_{mi}^{2}\sigma_{i,i}^2$ compared to Equation~(\ref{eq_A47}) describing the errors of the initial McNish-Lincoln predictions. However, as the initial McNish-Lincoln predictions use the smoothed radio flux delayed 6-months with respect to the current time, to estimate the accuracy, for instance, of the 1-month lead forecast, we need to calculate $\hat{\sigma}_{p}$ using Equation~(\ref{eq_A47}), which corresponds to 7-months ahead prediction.
At the same time, McNish-Lincoln+Kalman filter allows us to make the predictions without a 6-month delay, which in general reduces the prediction error $\hat{\sigma}_{p}$ calculated with Equation~(\ref{eq_A51}).


\end{document}